\documentclass[reprint,amsfonts,amsmath,amssymb,aps,pra,superscriptaddress,10pt]{revtex4-2}
\usepackage{graphicx}
\usepackage{sansmathfonts}
\usepackage[colorlinks=true,linkcolor=black,anchorcolor=black,citecolor=black,filecolor=black,menucolor=black,runcolor=black,urlcolor=black]{hyperref}

\makeatletter
\renewcommand*{\fnum@figure}{{\normalfont\bfseries \figurename~\thefigure}}
\renewcommand*{\@caption@fignum@sep}{ \textbar{} }
\makeatother
\let\oldcaption\caption
\renewcommand\caption[2][]{\oldcaption[#1]{\textbf{#1.} #2}}

\begin{document}

\title{Directional telecom photons from a chirally coupled quantum dot}

\author{Kristina Bodiroga}
\affiliation{Centre for Nanophotonics, Department of Physics, Engineering Physics \& Astronomy, 64 Bader Lane, Queen's University, Kingston, Ontario, Canada K7L 3N6}

\author{Jacob Ewaniuk}
\affiliation{Centre for Nanophotonics, Department of Physics, Engineering Physics \& Astronomy, 64 Bader Lane, Queen's University, Kingston, Ontario, Canada K7L 3N6}

\author{Andrew N. Wakileh}
\affiliation{Centre for Nanophotonics, Department of Physics, Engineering Physics \& Astronomy, 64 Bader Lane, Queen's University, Kingston, Ontario, Canada K7L 3N6}

\author{Lucas Rantz}
\affiliation{Centre for Nanophotonics, Department of Physics, Engineering Physics \& Astronomy, 64 Bader Lane, Queen's University, Kingston, Ontario, Canada K7L 3N6}

\author{Ivanna M. Boras Vazquez}
\affiliation{Centre for Nanophotonics, Department of Physics, Engineering Physics \& Astronomy, 64 Bader Lane, Queen's University, Kingston, Ontario, Canada K7L 3N6}

\author{Dan Dalacu}
\affiliation{National Research Council of Canada, Ottawa, Ontario, Canada K1A 0R6}
\affiliation{Centre for Nanophotonics, Department of Physics, Engineering Physics \& Astronomy, 64 Bader Lane, Queen's University, Kingston, Ontario, Canada K7L 3N6}

\author{Philip J. Poole}
\affiliation{National Research Council of Canada, Ottawa, Ontario, Canada K1A 0R6}

\author{Robin L. Williams}
\affiliation{National Research Council of Canada, Ottawa, Ontario, Canada K1A 0R6}

\author{Xiao-Liu Chu}
\affiliation{Institute for Digital Molecular Analytics and Science, Nanyang Technological University, Experimental Medicine Building, Level 6 \& 7, 59 Nanyang Drive, Singapore, 636921, Singapore}

\author{Nir Rotenberg}
\email{nir.rotenberg@queensu.ca}
\affiliation{Centre for Nanophotonics, Department of Physics, Engineering Physics \& Astronomy, 64 Bader Lane, Queen's University, Kingston, Ontario, Canada K7L 3N6}

\date{\today}

\begin{abstract}
Chiral quantum light–matter interfaces, where the internal spin state of a quantum emitter determines the direction in which it emits, are essential building blocks of non-reciprocal quantum devices, deterministic quantum logical gates and entanglement generation protocols. Yet, a chiral quantum interface that operates at telecom wavelengths, and is compatible with telecommunication infrastructure and silicon photonics, does not yet exist. Here, we report on an integrated chiral quantum interface in the original telecom band (1260-1360~nm), created by interfacing InAs quantum dots with a waveguide-coupled InP microdisk. We tune the quantum dot transitions through the photonic cavity using a strong magnetic field, observing a peak cavity enhancement of 3.3 and an emission directionality of 0.985, demonstrating the near-ideal chiral quantum coupling required for quantum information processing on integrated photonic devices.
\end{abstract}

\maketitle

\section{Introduction} \label{sec1}
The surprising discovery that quantum emitters coupled to photonic waveguides can be made to emit in a single direction, a scenario now known as chiral quantum optics \cite{Young:15,Lodahl:17}, redefined what is possible in quantum photonics. The directionality of these chiral interactions arises from the interplay of the phase of the transition dipole of the emitter and the local electric field of the photonic mode, effectively locking the emitter spin to a photon direction. This, in turn, has led to the realization of non-reciprocal few-photon devices \cite{Sollner:15,Scheucher:16} and deterministic quantum logical gates \cite{Shomroni:14,Schrinski:22}, largely realized with atomic systems. Chiral emission has also been observed in a variety of photonic structures coupled to quantum dots (QDs) \cite{Sollner:15, Coles:16, Barik:18, Mehrabad:20, Antoniadis:22, Mehrabad:23, Siampour:23, Rao:25, Mrowinski:19, Martin:25}, opening a route towards fully integrated solid-state quantum photonic devices based on these effects and motivating further proposals for quantum information processing devices \cite{Ralph:15,Sollner:15,Schrinski:22}, quantum networks \cite{Mahmoodian:16, Russo:18} and even neuromorphic quantum photonic processors \cite{Vazquez:26, Ewaniuk:25}.

\begin{figure*}[ht]
  \includegraphics[width=\textwidth]{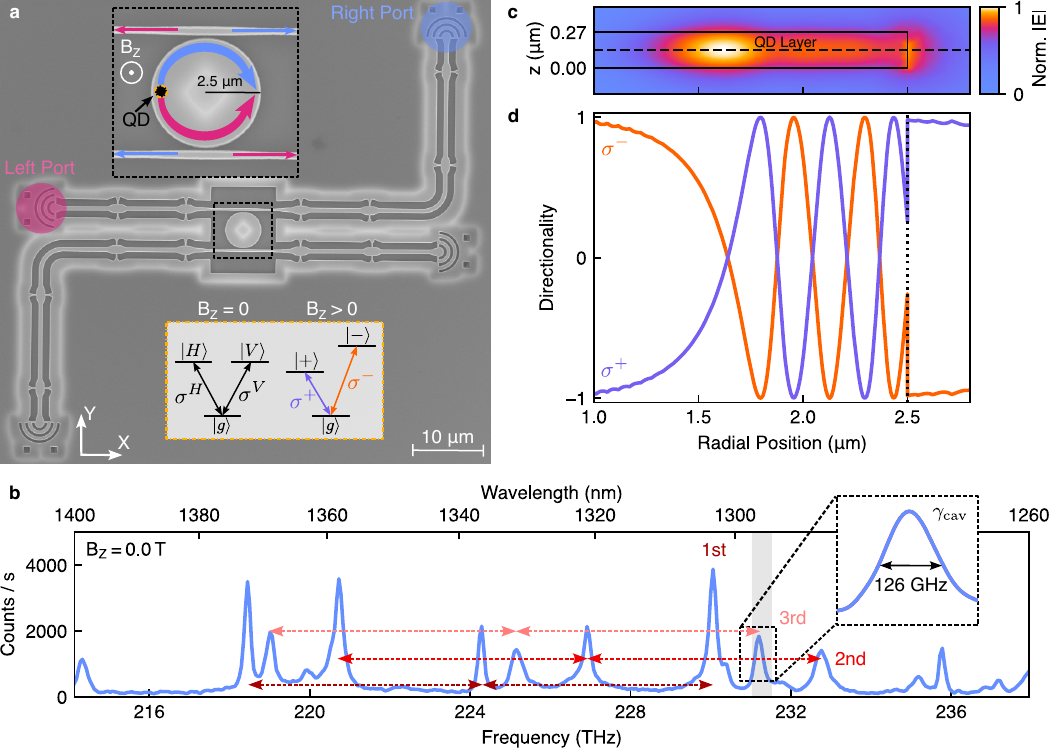}
  \caption[An integrated platform for chiral quantum photonics]{\textbf{a} Scanning electron microscope image of a 2.5~\textmu m radius-microdisk coupled to access waveguides with left (pink) and right (blue) ports for emission collection. The upper inset (dashed black line) shows a magnified view of the microdisk, where a QD near the edge is excited above-band. Under an applied out-of-plane magnetic field, $B_z$, QDs with circular transition dipoles couple to counter-propagating whispering-gallery-modes and are collected at opposite ports. The lower inset (orange dashed line) shows the energy levels of the relevant QD transition dipoles with no magnetic field (left), and when one is applied (right). \textbf{b} Above-band photoluminescence spectrum taken at 0~T across the telecom O-band, showing the first- (dark red), second- (red), and third- (light red) order cavity modes. Corresponding arrows indicate the free spectral range of each radial mode family. Our QD of interest lies spectrally near the third-order mode shown in the grey shaded region (inset in dashed black line magnifies the cavity mode, which has a measured width of $\gamma_\mathrm{cav}=126\pm7$~GHz). \textbf{c} Finite-element simulation showing the field amplitude of the cavity mode of interest in a cross-section of the microdisk. The plane containing the QDs is marked by a dashed black line. \textbf{d} Simulated directionality for the $\sigma^+$ and $\sigma^-$ transitions as a function of radial position in the microdisk for this third-order mode. The dotted black line denotes the outer edge of the disk. For near-ideal chiral coupling, a QD must be located near a region where $D=\pm 1$ and where there is significant field amplitude, within this boundary.}
  \label{fig:Fig1}
\end{figure*}

A key step towards emerging technologies, which we demonstrate here, is the observation of near-ideal chiral quantum interactions on integrated photonic devices at telecom-wavelengths. More specifically, we observe 98.5\% directional emission in the original telecom band from a self-assembled indium arsenide (InAs) QD into an integrated whispering-gallery-mode resonator, shown in Fig.~\ref{fig:Fig1}a.
We demonstrate this by using a magnetic field to tune a QD in and out of a photonic resonance, observing more than a three-fold reduction to the emitter lifetime when on resonance, while Zeeman-splitting of the transition energies makes the directional nature of the emission clearly visible in the output spectra. These results demonstrate a high-quality integrated chiral quantum element that is compatible with complex silicon photonic circuits and can provide critical functionalities in quantum information technologies.


\section{Quantum photonic platform} \label{sec2}
Microdisks, with whispering-gallery-mode resonances, can support quantum chiral light-matter interactions \cite{Brooks:21}. Here, we fabricate photonic circuits consisting of a microdisk with two access waveguides that connect to input and output couplers, as shown in Fig.~\ref{fig:Fig1}a, in an indium phosphide (InP) membrane with embedded QDs. We grow these QDs using chemical beam epitaxy in ultra-pure growth conditions, resulting in dots with excellent optical properties across both the telecom O- and C-bands at cryogenic temperatures \cite{Wakileh:26}. In our sample, the QDs are grown with a density of $\sim 50~\mathrm{QDs}/\text{\textmu m}^2$ and preferentially emit across the telecom O-band (1260-1360~nm).

While the waveguides that define our circuit are floating, the microdisks remain supported by central pedestals made of the indium gallium arsenide (InGaAs) sacrificial layer, which appear as dark regions in the centre of the disks (see upper inset to Fig.~\ref{fig:Fig1}a). We fabricate our disks with radii ranging from 2.5 to 4.5~\textmu m so that they support well-spaced resonances across the entire telecom band. This can be seen in the exemplary emission spectra of the QDs, where we excite the dots in a 2.5~\textmu m radius disk using above-band illumination, of which the O-band portion is presented in Fig.~\ref{fig:Fig1}b. In this spectrum, we observe a repeating set of 3 resonances, which we identify as the first-, second- and third-order radial modes by their free spectral range as noted on the figure. By fitting Lorentzian curves to all of the resonances, we find average quality factors (alongside 95\% confidence intervals) of $10149~[1078, 19220]$, $6304~[1885, 10723]$, and $3792~[688, 6897]$ for these respective modes (see Supplementary Sec.~S1 for more information).

In this work, we focus on the third-order resonance, whose relatively large spatial distribution assists in finding well-coupled QDs. As expected, the field maximum for this mode is located well inside the microdisk, as can be seen in Fig.~\ref{fig:Fig1}c, with two additional local maxima towards its edge. While the field is linearly polarized at the field maxima, the regions of circular polarization in between (and to the outside and inside) allow for chiral coupling. This is confirmed by the calculated emission directionality, $D$, as a function of the emitter radial position for this mode (see Supplementary Sec.~S2 for details), which we show in Fig.~\ref{fig:Fig1}d for the two circular dipoles, $\sigma^\pm$, corresponding to the transition dipoles of a neutral exciton in a QD in a strong, out-of-plane magnetic field $B_z$ (see lower inset to Fig.~\ref{fig:Fig1}a). The directionality ranges from -1 to 1, corresponding to perfect clockwise and counter-clockwise emission, respectively (see upper inset to Fig.~\ref{fig:Fig1}a), which in turn leads to output coupling at the left (pink) and right (blue) ports of the top access waveguide. Where $D=0$, for example near 1.6~\textmu m at the mode maximum, the emission is symmetric and equal in both directions. An ideal radial position for a chirally-coupled QD would therefore be in one of the four positions between 1.8 and 2.3~\textmu m, where one dipole has a directionality $D\approx+1$ and the other $D\approx-1$, and the field amplitude is still significant (and we note that this can be at any azimuthal angle along the disk).


\section{Telecom chiral quantum photonics} \label{sec3}
\begin{figure*}[ht]
  \centering
  \includegraphics[width=\textwidth]{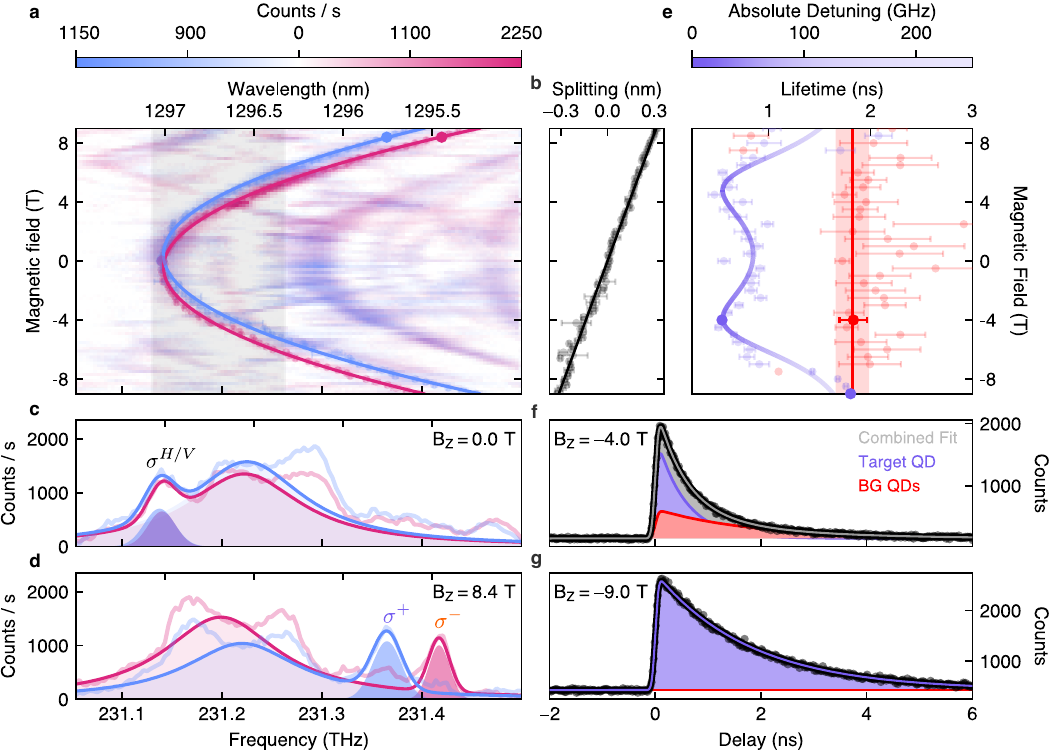}
  \caption[Chiral emission and enhancement into a photonic cavity]{\textbf{a} Emission spectra recorded from the left (pink) and right (blue) out-coupling ports as $B_z$ is varied from -9~T to 9~T. Symbols (error bars) show the emission frequencies (95\% confidence intervals) of each QD transition, which are overlaid with a fit (curves) as described in the main text. The darkened region highlights the resonant frequency and width of the cavity, and the enlargened markers correspond to \textbf{c,d}. \textbf{b} Measured Zeeman splitting as a function of the applied magnetic field (symbols with 95\% confidence interval error bars) and linear fit (solid curve). \textbf{c,d} Representative spectra measured at $B_z = 0$~T and $B_z = 8.4$~T, respectively, including fits to the cavity resonance and QD emission lines as explained in the text. \textbf{e} Lifetime of the $\sigma^+$ transition as a function of the applied magnetic field (purple points and curve), which changes as the transition shifts in and out of the cavity. The shade of the fit line indicates the absolute value of the cavity-QD detuning. Conversely, the lifetime of weakly coupled background (BG) QDs (shown in red) does not depend on the detuning. \textbf{f,g} Representative time-resolved lifetime measurements acquired at $B_z = -4.0$~T and $B_z = -9.0$~T, when the transition was on resonance (3~GHz, $0.02\gamma_\mathrm{cav}$ detuning) and detuned by 249~GHz ($1.98\gamma_{\mathrm{cav}}$), respectively. The fit (curve) to the data (symbols) contains a strong component corresponding to the target QD (purple shaded region) and a weaker, long-lived component from the BG QDs (red shaded region).}
  \label{fig:Fig2}
\end{figure*}

We now focus on the interaction of a single QD with a single resonance (see inset of Fig.~\ref{fig:Fig1}b). The QD is excited non-resonantly at low power (7~\textmu W at the cryostat entrance window, a saturation measurement is provided in Supplementary Sec.~S3), with photons collected from both output couplers for a series of out-of-plane magnetic fields ranging from -9~T to 9~T. We overlay the spectra collected from the left (pink) and right (blue) in Fig.~\ref{fig:Fig2}a,
subtracting out the microdisk resonance so as to distinguish the QDs more clearly, and noting that the spectral filter through which we measure has a 18~GHz bandwidth (see Supplementary Sec.~S4 for further details). While we observe many different QD transitions that shift with the magnetic field in this map, we focus on a single QD, marking the peaks of its resonances with red and blue symbols, corresponding to the emission direction. As is clearly evident, these transitions overlap at $B_z=0$, yet shift to higher frequencies and diverge as the amplitude of $B_z$ increases. We fit this magnetic field dependence of the emission frequencies of each dipole, $f_\pm\left(B_z\right)$, according to~\cite{Walck:98},
\begin{equation} \label{eq:Bdependence}
    hf_{\pm}\left(B_z\right) = hf_0 + \delta B_z^2 \pm \frac{1}{2}g\mu_BB_z,
\end{equation}
finding a natural frequency $f_0 = 231.1422\pm 0.0009$~THz (corresponding to 1297~nm), a diamagnetic shift $\delta = 14.67 \pm 0.01$~$\mathrm{\mu eV/T^2}$, and an exciton g-factor $g = -0.44\pm 0.02$ (as $\mu_B$ is the Bohr magneton, and the sign matches previous studies of similar QDs \cite{Kleemans:09}). These values are consistent with those for the neutral exciton of telecom-wavelength QDs \cite{Mikulicz:22, Sapienza:16}, which tend to be larger and so more weakly confine the electron and hole wavefunctions leading to both enhanced heavy-hole/light-hole coupling and to magnetic field-induced mixing of heavy-hole and light-hole states \cite{Tsai:08, Jovanov:12}. The Zeeman splitting of the two transitions displays the expected linear dependence on $B_z$, as can be more clearly seen in Fig.~\ref{fig:Fig2}b, where we show the measured splitting (symbols) and the curve corresponding to the linear term in Eq.~\ref{eq:Bdependence} (solid line).

For clarity, we show individual spectra, taken at $B_z=0$~T and $B_z=8.4$~T in Figs.~\ref{fig:Fig2}c and d, respectively. In these, we show the measured spectra (thick, lighter curves), which are overlaid with fits (thin, darker curves) that include the Lorentzian cavity (light shaded regions) and Gaussian emission lines for each dipole of the QD of interest (dark shaded blue and pink regions). Here, we more clearly see that at $B_z=0$~T the QD has a lower transition frequency than the cavity resonance, and an equal number of photons are found at the left and right output ports. In contrast, when $B_z=8.4$~T, the two transitions have split and are largely found in separate ports, and both have shifted to the higher-energy tail of the resonance. In each case, we find 6 additional QDs, which also tune with $B_z$, and are treated alongside the main QD while fitting (see Supplementary Sec.~S4 for full analysis details).

We observe the effect of the cavity on the QD by measuring its lifetime (from the right output port), presenting the results in Fig.~\ref{fig:Fig2}e. When the QD is on resonance with the cavity, at $B_z = -4.0$~T and 5.0~T, we find lifetimes down to $0.54\pm0.03$~ns and $0.61\pm0.07$~ns (where the errors are 95\% confidence intervals), respectively. In both cases, and indeed for the majority of applied $B_z$, we find this lifetime by fitting a bi-exponential curve to the measured data (see Supplementary Sec.~S5 for full details on lifetime measurement analysis), as shown in the exemplary curve for $B_z=-4.0$~T in Fig.~\ref{fig:Fig2}f. Here, the shorter lifetime (shaded purple region) corresponds to that of our target QD, while the weak (i.e. low-amplitude), longer-lived component (shaded red region) is the measured lifetime of weakly coupled background QDs that nevertheless radiate into the cavity ($\tau_\mathrm{BG} = 1.76 \pm 0.03$~ns). In contrast, when the QD transition is on the tail of the cavity resonance, at $B = -9.0$~T when it is detuned by 249~GHz (1.98$\gamma_\mathrm{cav}$), we find $\tau_\mathrm{QD} = 1.80 \pm 0.02$~ns (Fig.~\ref{fig:Fig2}g), approaching the off-resonant lifetime of the QD, $\tau_\mathrm{QD,0} = 1.93 \pm 0.13$~ns, which is extracted from the fit to the data and consistent with literature \cite{Wakileh:26}. By comparing the lifetimes of the QD when it is on- and off-resonance with the cavity, at -4.0~T (Fig.~\ref{fig:Fig2}f) and -9.0~T (Fig.~\ref{fig:Fig2}g) respectively, we find a peak measured emission enhancement of $3.3 \pm 0.4$ that, we note, is sufficient to overcome residual noise for gated QDs in, for example, slow-light photonic crystal waveguides \cite{Uppu:20, Albrechtsen:26}.

Following the most common definitions in the literature \cite{Mehrabad:20, Coles:16}, we calculate the directional contrast of the QD emission, $C_\pm$, for each dipole $\sigma^{\pm}$, by comparing the measured counts at the emission frequency of one dipole with those at the frequency of the orthogonal dipole, in the same output port. This is distinct from the directionality, $D_\pm$, where the emission from one dipole is compared across the two output ports, yet it is insensitive to imbalanced collection efficiencies from the two outputs, as described by S\"ollner \textit{et al.} \cite{Sollner:15} (see Supplementary Sec.~S6 for a full description of these quantities and their calculation). For example, at $B_z = 0$~T (Fig.~\ref{fig:Fig2}b), the transitions overlap and we calculate $C=0.035$ (averaged over $C_+$ and $C_-$) with a 95\% confidence interval $\left[0.000, 0.096\right]$. That is, in this case we observe symmetric emission, as expected for linear transition dipoles. Conversely, when $B_z=8.4$~T (Fig.~\ref{fig:Fig2}c), we measure $C=0.994~\left[0.957, 1.000\right]$, meaning that $\sigma^+$ and $\sigma^-$ radiate almost exclusively to counter-propagating modes and hence the QD is almost perfectly chirally-coupled to the photonic resonator. 

\begin{figure}[ht]
  \begin{center}
    \includegraphics[width=\columnwidth]{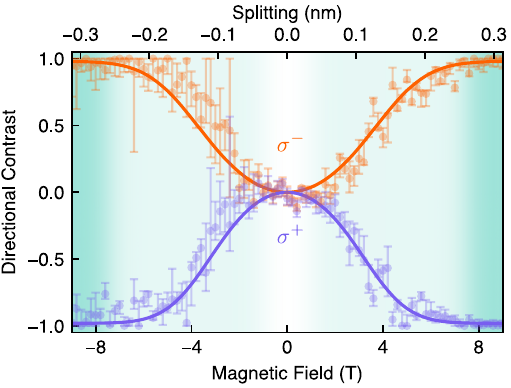}
  \end{center}
    \caption[Directional contrast for each QD transition dipole]{The directional contrast, $C_\pm$ (markers), and corresponding 95\% confidence intervals (error bars) as a function of the applied magnetic field, $B_z$, for each transition dipole $\sigma^\pm$ (purple and orange, respectively), calculated from the emission spectra of Fig.~\ref{fig:Fig2}a. Overlaying curves are a fit obtained by following the splitting (top axis, in correspondence with Fig.~\ref{fig:Fig2}d) of the Gaussian emission lines from each dipole (see Supplementary Sec.~S6). Shaded regions serve as a visual aid, denoting the fields at which the emission is symmetric (white), and where directional emission cannot be fully spectrally resolved (light teal) as opposed to where the transitions have sufficiently split (dark teal).}
  \label{fig:directionality}
\end{figure}

We plot the full, measured $B_z$-dependence of the directional contrast in Fig.~\ref{fig:directionality},
with the measured data (95\% confidence intervals) shown by symbols (error bars) with accompanying fits in solid lines. In this figure, we observe three distinct regions. First, within $\sim 1$~T of $B_z=0$~T (white region) we observe symmetric emission, as described above. Second, for large applied magnetic fields, $\left|B_z\right| \ge 7.5$~T (dark teal shaded region), we observe near-ideal chiral emission, with $C\rightarrow\pm 1$ for the $\sigma^\mp$ transitions. In between, we find regions of intermediate directional contrast (light teal shaded region), which arise because we are not able to spectrally resolve the different transitions. This is reflected in the fitted curves (solid lines), which use the measured Zeeman splitting (Fig.~\ref{fig:Fig2}d) and fit the QD transition linewidth as well as the directional contrast at high-amplitude fields (see Supplementary Sec.~S6 for details). From these, we find an overall directional contrast of $C=0.980~\left[0.939,0.994\right]$ for this QD-microdisk system. Additionally, we evaluate the overall directionality by calculating it from the measured counts only at the fields where the Zeeman splitting has sufficiently separated the dipoles (given the spectral resolution; dark teal shaded region in Fig.~\ref{fig:directionality}), and fitting the resultant data points, finding $D=0.985~\left[0.905, 0.998\right]$.


\section{Discussion} \label{sec:sec4}
\begin{figure}[ht!]
  \begin{center}
    \includegraphics[width=\columnwidth]{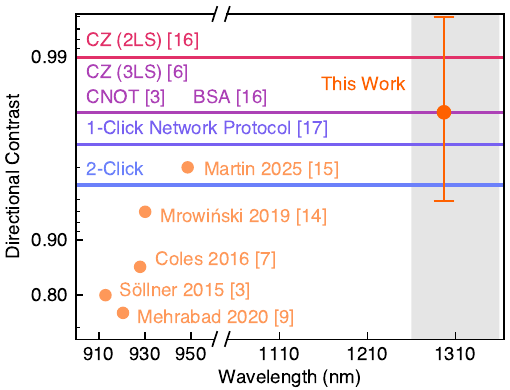}
  \end{center}
    \caption[Chiral quantum interfaces in the landscape of chiral quantum information protocols]{Coloured horizontal lines mark directional contrast thresholds required to achieve a success probability of 95\% (or 0.92 fidelity) for quantum information applications including 1- and 2-click quantum network protocols, Bell-state analyzers (BSAs), controlled-NOT (CNOT) gates, controlled-Z (CZ) gates with two- (2LS) and three-level systems (3LS). The circular markers show measured directional contrasts for QDs in a variety of nanophotonic structures that operate at different wavelengths. That from this work, which is the only in the O-band (shaded grey region), includes the 95\% confidence interval (error bars).}
  \label{fig:QCcontext}
\end{figure}

The promise of directional emission and quantum light-matter interactions in integrated photonic chips has stimulated the development of a large body of protocols for the generation of quantum resources and operations for quantum information processing. Inevitably, the success probability or fidelity of each protocol depends on the directionality of the quantum interface. In Fig.~\ref{fig:QCcontext},
we present a selection of protocols for critical functionalities, marking the directionality for each to reach its target performance threshold ($95\%$ success probability or $0.92$ fidelity, as indicated by the horizontal lines; see Supplementary Sec.~S6 for details). To be consistent with the rest of the manuscript, we assume that within the waveguide, prior to the output couplers, the directional contrast is equal to the directionality ($C=D$), a more commonly used figure of merit for protocols in the literature. The protocols considered include the use of a chiral quantum interface to realize a photonic controlled-NOT gate ($D=0.98$) \cite{Sollner:15}, a Bell-state analyzer ($D=0.98$) \cite{Ralph:15}, a controlled-Z gate based on two- ($D=0.99$) \cite{Ralph:15} and three-level systems ($D=0.98$) \cite{Schrinski:22}, and finally as a basis for a quantum network architecture ($D=0.95$ and $D=0.97$ for one-click and two-click protocols respectively) \cite{Mahmoodian:16}. 

In symbols, we show the measured directional contrasts that have been reported with QDs coupled to integrated photonic structures, as a function of their respective emission wavelengths, noting that there have been several other qualitative reports of chiral emission with these systems \cite{Barik:18, Antoniadis:22, Siampour:23}. All told, previous works have explored a breadth of nanophotonic structures, from broken-symmetry \cite{Sollner:15} and topological \cite{Mehrabad:20} photonic crystal waveguides, to strip waveguides \cite{Coles:16, Mrowinski:19} and more recently both ring resonators \cite{Rao:25} and photonic crystal cavities \cite{Martin:25}, at wavelengths ranging from 910-950~nm and with a peak reported $C=0.96$ \cite{Martin:25}. Here, we extend this to the telecom O-band, in an integrated photonic circuit at a wavelength near 1295 nm, and measure $C=0.98$, reaching the thresholds for all but one of the listed protocols. We note that we have identified other highly directional ($C=0.965,~0.970$) QDs in our system, along with a few that emit symmetrically ($C=0.053,~0.010$), which are shown in Supplementary Sec.~S7. Lastly, while the platform presented here operates in the telecom O-band, where zero-dispersion fibers are particularly attractive for short-range networks or within a large quantum computing architecture as envisioned by industry leaders \cite{Aghaee-Rad:25, Alexander:25}, the same QDs (in the same membranes) can be grown to emit in the C-band \cite{Wakileh:26} that, in conjunction with advances in heterogeneous integration \cite{Davanco:17, Larocque:24, Katsumi:23, Wang:26}, present a clear next step towards chiral quantum devices that are also compatible with long-range communications infrastructure.


\section{Methods} \label{sec:methods}

\noindent\textbf{Sample Growth and Fabrication}\\
The sample used in this work was grown via chemical beam epitaxy on a semi-insulating InP:Fe (001) substrate using trimethylindium, pre-cracked phosphine, and pre-cracked arsine as the indium, phosphorus, and arsenic sources, respectively. The native oxide was first removed at 580~$^\circ$C, followed by the growth of a 200~nm InP buffer layer at 530~$^\circ$C and a 1.2~\textmu m InGaAs sacrificial layer. A 270~nm-thick InP membrane containing a layer of QDs at its center was then grown, with growth conditions comparable to those of the high-density QD sample described in Ref.~\cite{Wakileh:26}. 

For device fabrication, a 120~nm silica (SiO\textsubscript{2}) hard mask was deposited by electron-beam physical vapour deposition. The structures were then patterned by electron-beam lithography using a 30~kV system and 275~nm of ZEP 520A resist. The pattern was aligned to the $[110]/[\bar{1}10]$ crystallographic directions, as these are the expected directions of the QD emission dipoles in this material. After resist development, the SiO\textsubscript{2} hard mask was dry etched by reactive ion etching (RIE) using an SF\textsubscript{6}:O\textsubscript{2} gas mixture. Any remaining resist was stripped in N-methyl-2-pyrrolidone at 75~$^{\circ}$C. The InP membrane was then etched by RIE using a Cl\textsubscript{2}:Ar gas mixture under the same power conditions. The remaining SiO\textsubscript{2} hard mask was removed using buffered oxide etchant, and the membrane was released by selectively wet etching the InGaAs sacrificial layer in a dilute piranha solution.\\

\noindent\textbf{Microdisk Quality Factor Measurements}\\
Spectrally resolved photoluminescence measurements were performed at 4~K using a closed-cycle helium cryostat. Quantum dots were excited via the wetting layer using continuous-wave excitation at a wavelength of 970~nm through a 100x cryogenic objective (0.81 numerical aperture). The excitation spot (1~\textmu m diameter) was positioned on the microdisk at high power to excite the QD ensemble within the disk well beyond ($>10$x) their saturation power. Given the high QD density, there was sufficient emission to fill the cavity modes across the telecom O-band. This was collected through the same objective, coupled into an SMF-28 fiber, and directed to a spectrometer equipped with a 0.4~nm-resolution grating and a liquid nitrogen-cooled InGaAs linear array detector.\\

\noindent\textbf{Quantum Dot Emission Spectra Measurements}\\
To obtain high-resolution spectra of the microdisk cavity resonance near the QD of interest, the QDs were excited via the wetting layer (970~nm, continuous-wave) at a reduced power, 7~\textmu W at the cryostat entrance window (16\% of the saturation power, see Supplementary Sec.~S3). The collected QD emission from each output port was directed to a $2\times 1$ fiber switch (enabling switching between emission directions), followed by a Gaussian-shaped tunable bandpass filter with 0.1~nm bandwidth. The center wavelength of the filter was swept over the wavelength range of interest in 0.02~nm steps. At each wavelength step, the filtered emission was directed to a superconducting nanowire single-photon detector, allowing the spectrum to be reconstructed from the wavelength-dependent photon counts. This measurement was performed for magnetic fields applied along the growth direction from -9~T to 9~T in steps of 0.2~T.\\

\noindent\textbf{Quantum Dot Emission Lifetime Measurements}\\
For lifetime measurements, 50~ps optical pulses were prepared by modulating a 970~nm continuous-wave laser using an Exail NIR-MX950-LN-20 amplitude modulator driven by a Tektronix AWG70001B arbitrary waveform generator, at a 10~MHz repetition rate. The pulses excite the QD above-band, and time differences between the pulse trigger signal and the detected photons were recorded. These measurements were repeated at applied magnetic fields ranging from -9~T to 9~T in steps of 0.5~T.


\section{Data Availability} \label{sec:data}
All data are available in the main text or Supplementary Information.
Correspondence and requests for materials should be addressed to
the corresponding authors (K.B., J.E. and N.R.).


\section{Acknowledgments} \label{sec:acknowledgments}
The authors thank G. Gibson for insightful discussions on fabrication, and gratefully acknowledge the support from the National Research Council of Canada (NRC), the Canadian Foundation for Innovation (CFI), the Ontario Ministry of Colleges, Universities, Research Excellence and Security (MCURES), the Natural Sciences and Engineering Research Council of Canada (NSERC), and Queen's University. X.-L.C. gratefully acknowledges funding support from Mitacs through the Globalink Research Award (Project No. IT46267).


\section{Author Contributions} \label{sec:contributions}
K.B. performed all numerical simulations and fabrication of the microdisk resonator structures on a wafer with QDs grown by A.W., D.D, P.P. and R.W. J.E. and X.-L.C. coordinated the experiment, under the supervision of N.R. QD emission spectra at varying magnetic fields were primarily measured by X.-L.C., with support from J.E. and I.B.V. Microdisk quality factors were measured by A.W. and K.B, and analyzed by K.B. Lifetime measurements were taken by J.E., I.B.V, A.W. and L.R., with all analysis conducted by L.R. Supplementary measurements were taken and analyzed by J.E. and A.W. Analysis of directional contrast and directionality was conducted by J.E., while the comparison with previous works was formed by X.-L.C. X.-L.C, K.B. and N.R wrote the manuscript with input from all of the authors.

\bibliography{references}

@article{Young:15,
title   = {{Polarization Engineering in Photonic Crystal Waveguides for Spin-Photon Entanglers}},
author  = {Young, A. B. and Thijssen, A. C. T. and Beggs, D. M. and Androvitsaneas, P. and Kuipers, L. and Rarity, J. G. and Hughes, S. and Oulton, R.},
journal = {Phys. Rev. Lett.},
volume  = {115},
number  = {15},
pages   = {153901},
year    = {2015},
doi     = {10.1103/PhysRevLett.115.153901},
}

@article{Walck:98,
title   = {{Exciton diamagnetic shift in semiconductor nanostructures}},
author  = {Walck, S. N. and Reinecke, T. L.},
journal = {Phys. Rev. B},
volume  = {57},
number  = {15},
pages   = {9088-9096},
year    = {1998},
doi     = {10.1103/PhysRevB.57.9088},
}

@article{Wakileh:26,
title   = {{Approaching Transform-Limited Line Widths in Telecom-Wavelength Transitions of Ungated Quantum Dots}},
author  = {Wakileh, Andrew N. and Dalacu, Dan and Poole, Philip J. and Lamontagne, Boris and Moisa, Simona and Williams, Robin L. and Rotenberg, Nir},
journal = {ACS Photonics},
volume  = {13},
number  = {6},
pages   = {1591-1598},
year    = {2026},
doi     = {10.1021/acsphotonics.5c02178},
}

@article{Uppu:20,
title   = {{Scalable integrated single-photon source}},
author  = {Ravitej Uppu and Freja T. Pedersen and Ying Wang and Cecilie T. Olesen and Camille Papon and Xiaoyan Zhou and Leonardo Midolo and Sven Scholz and Andreas D. Wieck and Arne Ludwig and Peter Lodahl},
journal = {Science Advances},
volume  = {6},
number  = {50},
pages   = {eabc8268},
year    = {2020},
doi     = {10.1126/sciadv.abc8268},
}

@article{Tsai:08,
title   = {{Diamagnetic Response of Exciton Complexes in Semiconductor Quantum Dots}},
author  = {Tsai, Ming-Fu and Lin, Hsuan and Lin, Chia-Hsien and Lin, Sheng-Di and Wang, Sheng-Yun and Lo, Ming-Cheng and Cheng, Shun-Jen and Lee, Ming-Chih and Chang, Wen-Hao},
journal = {Phys. Rev. Lett.},
volume  = {101},
number  = {26},
pages   = {267402},
year    = {2008},
doi     = {10.1103/PhysRevLett.101.267402},
}

@article{Sollner:15,
title   = {{Deterministic photon–emitter coupling in chiral photonic circuits}},
author  = {Söllner, Immo and Mahmoodian, Sahand and Hansen, Sofie Lindskov and Midolo, Leonardo and Javadi, Alisa and Kiršanskė, Gabija and Pregnolato, Tommaso and El-Ella, Haitham and Lee, Eun Hye and Song, Jin Dong and Stobbe, Søren and Lodahl, Peter},
journal = {Nature Nanotechnology},
volume  = {10},
number  = {9},
pages   = {775-778},
year    = {2015},
doi     = {10.1038/nnano.2015.159},
}

@article{Siampour:23,
title   = {{Observation of large spontaneous emission rate enhancement of quantum dots in a broken-symmetry slow-light waveguide}},
author  = {Siampour, Hamidreza and O’Rourke, Christopher and Brash, Alistair J. and Makhonin, Maxim N. and Dost, René and Hallett, Dominic J. and Clarke, Edmund and Patil, Pallavi K. and Skolnick, Maurice S. and Fox, A. Mark},
journal = {npj Quantum Information},
volume  = {9},
number  = {1},
pages   = {15},
year    = {2023},
doi     = {10.1038/s41534-023-00686-9},
}

@article{Scheucher:16,
title   = {{Quantum optical circulator controlled by a single chirally coupled atom}},
author  = {Michael Scheucher and Adèle Hilico and Elisa Will and Jürgen Volz and Arno Rauschenbeutel},
journal = {Science},
volume  = {354},
number  = {6319},
pages   = {1577-1580},
year    = {2016},
doi     = {10.1126/science.aaj2118},
}

@article{Sapienza:16,
title   = {{Magneto-optical spectroscopy of single charge-tunable InAs/GaAs quantum dots emitting at telecom wavelengths}},
author  = {Sapienza, Luca and Al-Khuzheyri, Rima and Dada, Adetunmise and Griffiths, Andrew and Clarke, Edmund and Gerardot, Brian D.},
journal = {Phys. Rev. B},
volume  = {93},
number  = {15},
pages   = {155301},
year    = {2016},
doi     = {10.1103/PhysRevB.93.155301},
}

@article{Ralph:15,
title   = {{Photon Sorting, Efficient Bell Measurements, and a Deterministic Controlled-$Z$ Gate Using a Passive Two-Level Nonlinearity}},
author  = {Ralph, T. C. and S\"ollner, I. and Mahmoodian, S. and White, A. G. and Lodahl, P.},
journal = {Phys. Rev. Lett.},
volume  = {114},
number  = {17},
pages   = {173603},
year    = {2015},
doi     = {10.1103/PhysRevLett.114.173603},
}

@article{Mikulicz:22,
title   = {{Diamagnetic coefficients and $g$-factors of InAs/InGaAlAs quantum dashes emitting at telecom wavelengths}},
author  = {Mikulicz, M. G. and Mrowiński, P. and Gawełczyk, M. and Reithmaier, J. P. and Höfling, S. and Sęk, G.},
journal = {Journal of Applied Physics},
volume  = {132},
number  = {14},
pages   = {144301},
year    = {2022},
doi     = {10.1063/5.0101345},
}

@article{Mehrabad:23,
title   = {{Chiral topological add--drop filter for integrated quantum photonic circuits}},
author  = {M. Jalali Mehrabad and A. P. Foster and N. J. Martin and R. Dost and E. Clarke and P. K. Patil and M. S. Skolnick and L. R. Wilson},
journal = {Optica},
volume  = {10},
number  = {3},
pages   = {415-421},
year    = {2023},
doi     = {10.1364/OPTICA.481684},
}

@article{Mehrabad:20,
title   = {{Chiral topological photonics with an embedded quantum emitter}},
author  = {Mahmoud Jalali Mehrabad and Andrew P. Foster and Ren\'{e} Dost and Edmund Clarke and Pallavi K. Patil and A. Mark Fox and Maurice S. Skolnick and Luke R. Wilson},
journal = {Optica},
volume  = {7},
number  = {12},
pages   = {1690-1696},
year    = {2020},
doi     = {10.1364/OPTICA.393035},
}

@article{Martin-Cano:19,
title   = {{Chiral Emission into Nanophotonic Resonators}},
author  = {Martin-Cano, Diego and Haakh, Harald R. and Rotenberg, Nir},
journal = {ACS Photonics},
volume  = {6},
number  = {4},
pages   = {961-966},
year    = {2019},
doi     = {10.1021/acsphotonics.8b01555},
}

@article{Martin:25,
title   = {{Purcell-enhanced, directional light-matter interaction in a waveguide-coupled nanocavity}},
author  = {Nicholas J. Martin and Dominic Hallett and Mateusz Duda and Luke Hallacy and Elena Callus and Luke Brunswick and Ren\'{e} Dost and Edmund Clarke and Pallavi K. Patil and Pieter Kok and Maurice S. Skolnick and Luke R. Wilson},
journal = {Optica},
volume  = {12},
number  = {7},
pages   = {1100-1108},
year    = {2025},
doi     = {10.1364/OPTICA.561630},
}

@article{Mahmoodian:16,
title   = {{Quantum Networks with Chiral-Light--Matter Interaction in Waveguides}},
author  = {Mahmoodian, Sahand and Lodahl, Peter and S\o{}rensen, Anders S.},
journal = {Phys. Rev. Lett.},
volume  = {117},
number  = {24},
pages   = {240501},
year    = {2016},
doi     = {10.1103/PhysRevLett.117.240501},
}

@article{Lodahl:17,
title   = {{Chiral quantum optics}},
author  = {Lodahl, Peter and Mahmoodian, Sahand and Stobbe, Søren and Rauschenbeutel, Arno and Schneeweiss, Philipp and Volz, Jürgen and Pichler, Hannes and Zoller, Peter},
journal = {Nature},
volume  = {541},
number  = {7638},
pages   = {473-480},
year    = {2017},
doi     = {10.1038/nature21037},
}

@article{Kleemans:09,
title   = {{Size-dependent exciton $g$ factor in self-assembled InAs/InP quantum dots}},
author  = {Kleemans, N. A. J. M. and van Bree, J. and Bozkurt, M. and van Veldhoven, P. J. and Nouwens, P. A. and N\"otzel, R. and Silov, A. Yu. and Koenraad, P. M. and Flatt\'e, M. E.},
journal = {Phys. Rev. B},
volume  = {79},
number  = {4},
pages   = {045311},
year    = {2009},
doi     = {10.1103/PhysRevB.79.045311},
}

@article{Jovanov:12,
title   = {{Highly nonlinear excitonic Zeeman spin splitting in composition-engineered artificial atoms}},
author  = {Jovanov, V. and Eissfeller, T. and Kapfinger, S. and Clark, E. C. and Klotz, F. and Bichler, M. and Keizer, J. G. and Koenraad, P. M. and Brandt, M. S. and Abstreiter, G. and Finley, J. J.},
journal = {Phys. Rev. B},
volume  = {85},
number  = {16},
pages   = {165433},
year    = {2012},
doi     = {10.1103/PhysRevB.85.165433},
}

@article{Coles:16,
title   = {{Chirality of nanophotonic waveguide with embedded quantum emitter for unidirectional spin transfer}},
author  = {Coles, R. J. and Price, D. M. and Dixon, J. E. and Royall, B. and Clarke, E. and Kok, P. and Skolnick, M. S. and Fox, A. M. and Makhonin, M. N.},
journal = {Nature Communications},
volume  = {7},
number  = {1},
pages   = {11183},
year    = {2016},
doi     = {10.1038/ncomms11183},
}

@article{Brooks:21,
title   = {{Integrated Whispering-Gallery-Mode Resonator for Solid-State Coherent Quantum Photonics}},
author  = {Brooks, Arianne and Chu, Xiao-Liu and Liu, Zhe and Schott, Rüdiger and Ludwig, Arne and Wieck, Andreas D. and Midolo, Leonardo and Lodahl, Peter and Rotenberg, Nir},
journal = {Nano Letters},
volume  = {21},
number  = {20},
pages   = {8707-8714},
year    = {2021},
doi     = {10.1021/acs.nanolett.1c02818},
}

@misc{Vazquez:26,
title   = {{Quantum photonic neural networks in time}},
author  = {Ivanna M. Boras Vazquez and Jacob Ewaniuk and Nir Rotenberg},
year    = {2026},
howpublished = {Preprint at https://arxiv.org/abs/2603.23798},
}

@article{Barik:18,
title   = {{A topological quantum optics interface}},
author  = {Sabyasachi Barik and Aziz Karasahin and Christopher Flower and Tao Cai and Hirokazu Miyake and Wade DeGottardi and Mohammad Hafezi and Edo Waks},
journal = {Science},
volume  = {359},
number  = {6376},
pages   = {666-668},
year    = {2018},
doi     = {10.1126/science.aaq0327},
}

@article{Antoniadis:22,
title   = {{A chiral one-dimensional atom using a quantum dot in an open microcavity}},
author  = {Antoniadis, Nadia O. and Tomm, Natasha and Jakubczyk, Tomasz and Schott, Rüdiger and Valentin, Sascha R. and Wieck, Andreas D. and Ludwig, Arne and Warburton, Richard J. and Javadi, Alisa},
journal = {npj Quantum Information},
volume  = {8},
number  = {1},
pages   = {27},
year    = {2022},
doi     = {10.1038/s41534-022-00545-z},
}

@article{Albrechtsen:26,
title   = {{A quantum-coherent photon--emitter interface in the original telecom band}},
author  = {Albrechtsen, M. and Krüger, S. and Loredo, J. C. and Stefan, L. and Liu, Z. and Meng, Y. and Niekamp, L. L. and Seyschab, B. F. and Spitzer, N. and Warburton, R. J. and Lodahl, P. and Ludwig, A. and Midolo, L.},
journal = {Nature Nanotechnology},
volume  = {21},
pages   = {642-647},
year    = {2026},
doi     = {10.1038/s41565-026-02156-7},
}

@article{Schrinski:22,
title   = {{Passive Quantum Phase Gate for Photons Based on Three Level Emitters}},
author  = {B. Schrinski and M. Lamaison and A. S. Sørensen},
journal = {Phys. Rev. Lett.},
volume  = {129},
pages   = {130502},
year    = {2022},
}

@article{Shomroni:14,
title   = {{All-optical routing of single photons by a one-atom switch controlled by a single photon}},
author  = {I. Shomroni and S. Rosenblum and Y. Lovsky and O. Bechler and G. Guendelman and B. Dayan},
journal = {Science},
volume  = {345},
pages   = {903-906},
year    = {2014},
}

@article{Russo:18,
title   = {{Photonic graph state generation from quantum dots and color centers for quantum communications}},
author  = {Russo, Antonio and Barnes, Edwin and Economou, Sophia E.},
journal = {Phys. Rev. B},
volume  = {98},
number  = {8},
pages   = {085303},
year    = {2018},
doi     = {10.1103/PhysRevB.98.085303},
}

@misc{Ewaniuk:25,
title   = {{Large-Scale Tree-Type Photonic Cluster State Generation with Recurrent Quantum Photonic Neural Networks}},
author  = {Jacob Ewaniuk and Bhavin J. Shastri and Nir Rotenberg},
year    = {2025},
howpublished = {Preprint at https://arxiv.org/abs/2505.14628},
}

@article{Aghaee-Rad:25,
title   = {{Scaling and networking a modular photonic quantum computer}},
author  = {Aghaee Rad, H. and Ainsworth, T. and Alexander, R. N. and Altieri, B. and Askarani, M. F. and Baby, R. and Banchi, L. and Baragiola, B. Q. and Bourassa, J. E. and Chadwick, R. S. and Charania, I. and Chen, H. and Collins, M. J. and Contu, P. and D'Arcy, N. and Dauphinais, G. and De Prins, R. and Deschenes, D. and Di Luch, I. and Duque, S. and Edke, P. and Fayer, S. E. and Ferracin, S. and Ferretti, H. and Gefaell, J. and Glancy, S. and González-Arciniegas, C. and Grainge, T. and Han, Z. and Hastrup, J. and Helt, L. G. and Hillmann, T. and Hundal, J. and Izumi, S. and Jaeken, T. and Jonas, M. and Kocsis, S. and Krasnokutska, I. and Larsen, M. V. and Laskowski, P. and Laudenbach, F. and Lavoie, J. and Li, M. and Lomonte, E. and Lopetegui, C. E. and Luey, B. and Lund, A. P. and Ma, C. and Madsen, L. S. and Mahler, D. H. and Mantilla Calderón, L. and Menotti, M. and Miatto, F. M. and Morrison, B. and Nadkarni, P. J. and Nakamura, T. and Neuhaus, L. and Niu, Z. and Noro, R. and Papirov, K. and Pesah, A. and Phillips, D. S. and Plick, W. N. and Rogalsky, T. and Rortais, F. and Sabines-Chesterking, J. and Safavi-Bayat, S. and Sazhaev, E. and Seymour, M. and Rezaei Shad, K. and Silverman, M. and Srinivasan, S. A. and Stephan, M. and Tang, Q. Y. and Tasker, J. F. and Teo, Y. S. and Then, R. B. and Tremblay, J. E. and Tzitrin, I. and Vaidya, V. D. and Vasmer, M. and Vernon, Z. and Villalobos, L. F. S. S. M. and Walshe, B. W. and Weil, R. and Xin, X. and Yan, X. and Yao, Y. and Zamani Abnili, M. and Zhang, Y.},
journal = {Nature},
volume  = {638},
pages   = {912-919},
year    = {2025},
doi     = {10.1038/s41586-024-08406-9},
}

@article{Alexander:25,
title   = {{A manufacturable platform for photonic quantum computing}},
author  = {Alexander, Koen and Benyamini, Avishai and Black, Dylan and Bonneau, Damien and Burgos, Stanley and Burridge, Ben and Cable, Hugo and Campbell, Geoff and Catalano, Gabriel and Ceballos, Alejandro and Chang, Chia-Ming and Choudhury, Sourav Sen and Chung, C. J. and Danesh, Fariba and Dauer, Tom and Davis, Michael and Dudley, Eric and Er-Xuan, Ping and Fargas, Josep and Farsi, Alessandro and Fenrich, Colleen and Frazer, Jonathan and Fukami, Masaya and Ganesan, Yogeeswaran and Gibson, Gary and Gimeno-Segovia, Mercedes and Goeldi, Sebastian and Goley, Patrick and Haislmaier, Ryan and Halimi, Sami and Hansen, Paul and Hardy, Sam and Horng, Jason and House, Matthew and Hu, Hong and Jadidi, Mehdi and Jain, Vijay and Johansson, Henrik and Jones, Thomas and Kamineni, Vimal and Kelez, Nicholas and Koustuban, Ravi and Kovall, George and Krogen, Peter and Kumar, Nikhil and Liang, Yong and LiCausi, Nicholas and Llewellyn, Dan and Lokovic, Kimberly and Lovelady, Michael and Manfrinato, Vitor Riseti and Melnichuk, Ann and Mendoza, Gabriel and Moores, Brad and Mukherjee, Shaunak and Munns, Joseph and Musalem, Francois-Xavier and Najafi, Faraz and O'Brien, Jeremy L. and Ortmann, J. Elliott and Pai, Sunil and Park, Bryan and Peng, Hsuan-Tung and Penthorn, Nicholas and Peterson, Brennan and Peterson, Gabriel and Poush, Matt and Pryde, Geoff J. and Ramprasad, Tarun and Ray, Gareth and Rodriguez, Angelita Viejo and Roxworthy, Brian and Rudolph, Terry and Saunders, Dylan J. and Shadbolt, Pete and Shah, Deesha and Bahgat Shehata, Andrea and Shin, Hyungki and Sinsky, Jeffrey and Smith, Jake and Sohn, Ben and Sohn, Young-Ik and Son, Gyeongho and Souza, Mario C. M. M. and Sparrow, Chris and Staffaroni, Matteo and Stavrakas, Camille and Sukumaran, Vijay and Tamborini, Davide and Thompson, Mark G. and Tran, Khanh and Triplett, Mark and Tung, Maryann and Veitia, Andrzej and Vert, Alexey and Vidrighin, Mihai D. and Vorobeichik, Ilya and Weigel, Peter and Wingert, Matthew and Wooding, Jamie and Zhou, Xinran and PsiQuantum team},
journal = {Nature},
volume  = {641},
pages   = {876-883},
year    = {2025},
doi     = {10.1038/s41586-025-08820-7},
}

@article{Mrowinski:19,
title   = {{Directional Emission of a Deterministically Fabricated Quantum Dot--Bragg Reflection Multimode Waveguide System}},
author  = {Mrowiński, Paweł and Schnauber, Peter and Gutsche, Philipp and Kaganskiy, Arsenty and Schall, Johannes and Burger, Sven and Rodt, Sven and Reitzenstein, Stephan},
journal = {ACS Photonics},
volume  = {6},
number  = {9},
pages   = {2231-2237},
year    = {2019},
doi     = {10.1021/acsphotonics.9b00369},
}

@article{Rao:25,
title   = {{Chiral Single Photon Routing via Cavity-Assisted Spin-Momentum Locking}},
author  = {Rao, Mujie and Yang, Jiawei and Song, Changkun and Wang, Yangpeng and Wang, Feiyue and Liu, Shunfa and Bie, Yaqing and Liu, Jin and Yu, Ying and Yu, Siyuan},
journal = {Nano Letters},
volume  = {25},
number  = {29},
pages   = {11406-11412},
year    = {2025},
doi     = {10.1021/acs.nanolett.5c02598},
}

@article{Davanco:17,
title   = {{Heterogeneous integration for on-chip quantum photonic circuits with single quantum dot devices}},
author  = {Marcelo Davanco and Jin Liu and Luca Sapienza and Chen-Zhao Zhang and José Vinícius De Miranda Cardoso and Varun Verma and Richard Mirin and Sae Woo Nam and Liu Liu and Kartik Srinivasan},
journal = {Nature Communications},
volume  = {8},
number  = {889},
year    = {2017},
doi     = {10.1038/s41467-017-00987-6},
}

@article{Larocque:24,
title   = {{Tunable quantum emitters on large-scale foundry silicon photonics}},
author  = {Hugo Larocque and Mustafa Atabey Buyukkaya and Carlos Errando-Herranz and Camille Papon and Samuel Harper and Max Tao and Jacques Carolan and Chang-Min Lee and Christopher J. K. Richardson and Gerald L. Leake and Daniel J. Coleman and Michael L. Fanto and Edo Waks and Dirk Englund},
journal = {Nature Communications},
volume  = {15},
number  = {5781},
year    = {2024},
doi     = {10.1038/s41467-024-50208-0},
}

@article{Wang:26,
title   = {{Scalable all-solid-state cavity QED on a hybrid quantum dot--lithium niobate platform}},
author  = {Xudong Wang and Yifan Zhu and Xiuqi Zhang, Yuanhao Qin and Yang Chen and Runze Liu and Junyi Zhao and Yiyang Lou and Yongheng Huo and Xin Ou and Jiaxiang Zhang},
journal = {npj Quantum Information},
year    = {2026},
doi     = {10.1038/s41534-026-01282-3},
}

@article{Katsumi:23,
title   = {{CMOS-compatible integration of telecom band InAs/InP quantum-dot single-photon sources on a Si chip using transfer printing}},
author  = {Katsumi, Ryota and Ota, Yasutomo and Tajiri, Takeyoshi and Iwamoto, Satoshi and Ranbir, Kaur and Reithmaier, Johann Peter and Benyoucef, Mohamed and Arakawa, Yasuhiko},
journal = {Applied Physics Express},
volume  = {16},
number  = {1},
pages   = {012004},
year    = {2022},
doi     = {10.35848/1882-0786/acabaa},
}

@article{Barbour:11,
title   = {{A tunable microcavity}},
author  = {Barbour, Russell J. and Dalgarno, Paul A. and Curran, Arran and Nowak, Kris M. and Baker, Howard J. and Hall, Denis R. and Stoltz, Nick G. and Petroff, Pierre M. and Warburton, Richard J.},
journal = {Journal of Applied Physics},
volume  = {110},
number  = {5},
pages   = {053107},
year    = {2011},
doi     = {10.1063/1.3632057},
}

\end{document}


\title{Supplementary Information for\\``Directional telecom photons from a chirally coupled quantum dot''}

\author{Kristina Bodiroga}
\affiliation{Centre for Nanophotonics, Department of Physics, Engineering Physics \& Astronomy, 64 Bader Lane, Queen's University, Kingston, Ontario, Canada K7L 3N6}

\author{Jacob Ewaniuk}
\affiliation{Centre for Nanophotonics, Department of Physics, Engineering Physics \& Astronomy, 64 Bader Lane, Queen's University, Kingston, Ontario, Canada K7L 3N6}

\author{Andrew N. Wakileh}
\affiliation{Centre for Nanophotonics, Department of Physics, Engineering Physics \& Astronomy, 64 Bader Lane, Queen's University, Kingston, Ontario, Canada K7L 3N6}

\author{Lucas Rantz}
\affiliation{Centre for Nanophotonics, Department of Physics, Engineering Physics \& Astronomy, 64 Bader Lane, Queen's University, Kingston, Ontario, Canada K7L 3N6}

\author{Ivanna M. Boras Vazquez}
\affiliation{Centre for Nanophotonics, Department of Physics, Engineering Physics \& Astronomy, 64 Bader Lane, Queen's University, Kingston, Ontario, Canada K7L 3N6}

\author{Dan Dalacu}
\affiliation{National Research Council of Canada, Ottawa, Ontario, Canada K1A 0R6}
\affiliation{Centre for Nanophotonics, Department of Physics, Engineering Physics \& Astronomy, 64 Bader Lane, Queen's University, Kingston, Ontario, Canada K7L 3N6}

\author{Philip J. Poole}
\affiliation{National Research Council of Canada, Ottawa, Ontario, Canada K1A 0R6}

\author{Robin L. Williams}
\affiliation{National Research Council of Canada, Ottawa, Ontario, Canada K1A 0R6}

\author{Xiao-Liu Chu}
\affiliation{Institute for Digital Molecular Analytics and Science, Nanyang Technological University, Experimental Medicine Building, Level 6 \& 7, 59 Nanyang Drive, Singapore, 636921, Singapore}

\author{Nir Rotenberg}
\email{nir.rotenberg@queensu.ca}
\affiliation{Centre for Nanophotonics, Department of Physics, Engineering Physics \& Astronomy, 64 Bader Lane, Queen's University, Kingston, Ontario, Canada K7L 3N6}

\date{\today}

\maketitle

\section{Fitting Microdisk Quality Factors} \label{sec:Qfactors}
\begin{figure}[htb]
    \includegraphics[width=\columnwidth]{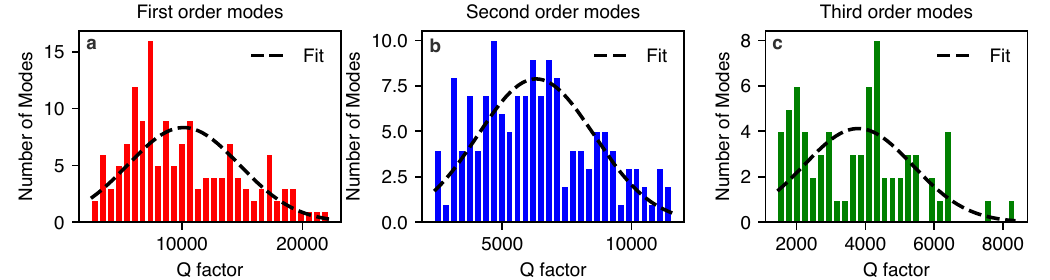}
    \caption[Distributions of microdisk cavity resonance quality factors by radial mode order]{Histograms of the measured quality factors for \textbf{a} first- (147 instances, red), \textbf{b} second- (145 instances, blue), and \textbf{c} third-order (71 instances, green) radial modes. The dashed black curves show fits to normal distributions, resulting in mean (with 95\% confidence intervals in brackets) quality factors of $10149~[1078, 19220]$, $6304~[1885, 10723]$, and $3792~[688, 6897]$ for the first-, second-, and third-order modes, respectively.}
    \label{fig:quality}
\end{figure}

Following the procedure described in Sec.~V, microdisk cavity resonances were resolved by using a high-power (well above saturation, see Sec.~\ref{sec:saturation} for more details) excitation laser to excite the QDs above-band. Spectra spanning 1260-1600~nm were acquired for 27 individual microdisks, and 363 cavity resonances were respectively fit using a Lorentzian line shape,
\begin{equation} \label{eq:cavity}
    I_\mathrm{cav}(f) = I_\mathrm{cav,0}\frac{\left(\frac{\gamma_\mathrm{cav}}{2}\right)^2}{(f-f_\mathrm{cav})^2+\left(\frac{\gamma_\mathrm{cav}}{2}\right)^2} + I_\mathrm{bg},
\end{equation}
where $I_\mathrm{cav,0}$ is the peak intensity, $\gamma_\mathrm{cav}$ is the cavity full width at half maximum (FWHM), $f_\mathrm{cav}$ is its resonance frequency, and $I_\mathrm{bg}$ is a constant background. From each fit, the quality factor is estimated as,
\begin{equation} \label{eq:quality}
    Q = \frac{f_\mathrm{cav}}{\gamma_\mathrm{cav}}.
\end{equation}
Subsequently, the radial mode order of each resonance was identified by the differences in free spectral range. This allowed us to group the quality factors by mode order, resulting in the histograms shown in Fig.~\ref{fig:quality}.
Each histogram was fit to a normal distribution, finding mean (with 95\% confidence intervals in brackets) quality factors of $10149~[1078, 19220]$, $6304~[1885, 10723]$, and $3792~[688, 6897]$ for the first-, second-, and third-order modes, respectively.

\section{Numerical Simulation of Microdisk Radial Modes and Directionality} \label{sec:comsol}
\begin{figure}[htb]
    \includegraphics[width=\columnwidth]{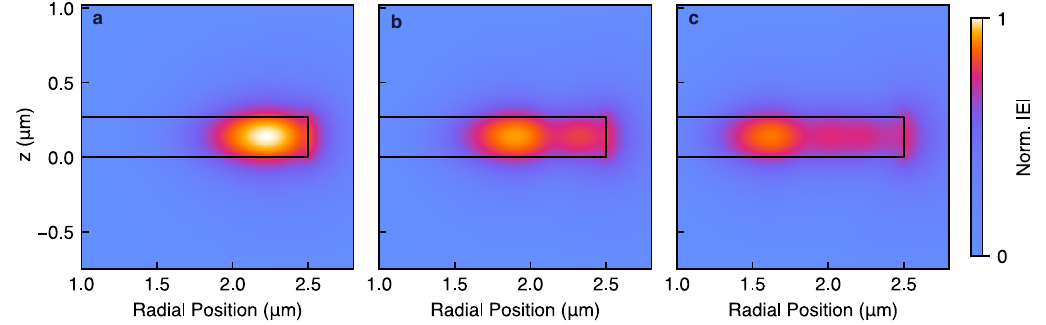}
    \caption[Simulated electric field intensity profiles]{Normalized electric field intensity distributions for the \textbf{a} first-, \textbf{b} second-, and \textbf{c} third-order radial modes. The colour scale is normalized to the maximum field intensity of the first-order mode. Simulated resonances occur at 229.5~THz ($m_\mathrm{az}$ = 27), 236.7~THz ($m_\mathrm{az}$ = 24), and 233.8~THz ($m_\mathrm{az}$ = 20) for the first-, second-, and third-order modes, respectively. These simulated modes correspond to the experimentally observed resonances shown in Fig. 1c, with frequency offsets attributed to fabrication-induced dimensional variations.}
    \label{fig:comsol}
\end{figure}

The microdisk resonator was modelled using a 2D axisymmetric geometry in COMSOL Multiphysics. The simulation consisted of an InP disk surrounded by an air domain encased with a perfectly matched layer (PML) to absorb outgoing waves and deduce reflections (see Fig.~\ref{fig:comsol}).
To ensure simulation convergence, the InP disk was meshed with a maximum element size equal to one fifth of the disk thickness, while the surrounding air domain was meshed with a maximum element size of one third of the wavelength. The air domain extended by one half wavelength beyond the disk in both the radial and $z$ directions, and was enclosed by an additional half-wavelength-thick PML. The modes were calculated using the \textit{Electromagnetic Waves, Frequency Domain} physics interface. Owing to the rotational symmetry of the structure, the azimuthal field dependence was represented by an out-of-plane wave number corresponding to the azimuthal mode number, $m_\mathrm{az}$. A parametric sweep of $m_\mathrm{az}$ was performed near the target resonance frequency of 231.2~THz to identify the supported mode families. For each azimuthal mode number, an eigenfrequency study was used to determine the resonant modes. The azimuthal mode number corresponding to the desired third-order radial mode ($m_\mathrm{az} = 20$) was identified, yielding a simulated resonance at 233.8~THz. The difference between the simulated and target resonance frequencies is attributed to fabrication-induced variations.

We follow the directionality calculation described in Ref.~\cite{Martin-Cano:19}. The effective mode volume for circularly polarized dipoles coupled to the counter-propagating whispering-gallery modes, $V_\pm(\mathbf{r})$, was calculated in cylindrical coordinates as,
\begin{equation}
V_{\pm}(\mathbf{r}) =
\frac{
\int d^3\mathbf{r} \left[
\varepsilon\left( -E_r^2 - E_z^2 + E_\phi^2 \right)
- \mu \left( H_r^2 + H_z^2 - H_\phi^2 \right)
\right]
}{
-\varepsilon\left[ E_r(\mathbf{r}) \mp i E_\phi(\mathbf{r}) \right]^2
},
\end{equation}
where $\mathbf{E}$ is the electric field, $\mathbf{H}$ is the magnetic field, $\varepsilon$ is the permittivity and $\mu$ is the permeability of the material. The position-dependent chiral Purcell factor, $F_\pm(\textbf{r})$, was then calculated from the effective mode volume,
\begin{equation}
    F_\pm(\textbf{r})=\frac{3}{4\pi^2}\bigg(\frac{\lambda_\mathrm{cav}}{n}\bigg)^3\Re{\bigg(\frac{Q}{V_\pm(\textbf{r})}\bigg)},
\end{equation}
where $n=3.17$ is the refractive index of InP, $\lambda_\mathrm{cav}$ is the resonant wavelength, and $Q=1883$ is the experimentally measured quality factor of the cavity resonance. The spatial distribution of $F_\pm(\textbf{r})$ describes the Purcell enhancement experienced by $\sigma^+$ and $\sigma^-$ circularly polarized dipoles, thereby identifying both the local handedness of the optical mode and the strength of the chiral light-matter interaction. The directionality was then determined from the relative Purcell enhancement of the two circular dipoles,
\begin{equation}
    D_\pm(\textbf{r}) = \frac{F_\pm(\textbf{r})-F_\mp(\textbf{r})}{F_+(\textbf{r})+F_-(\textbf{r})},
\end{equation}
where $D_+$ and $D_-$
correspond to the directionality for the $\sigma^+$ and $\sigma^-$
dipoles, respectively. The calculated directionality maps are shown in Fig.~1d of the main text and are in good agreement with the results reported in Ref.~\cite{Martin-Cano:19}.

\section{Saturation of the Quantum Dot Emission} \label{sec:saturation}
\begin{figure}[htb]
    \includegraphics{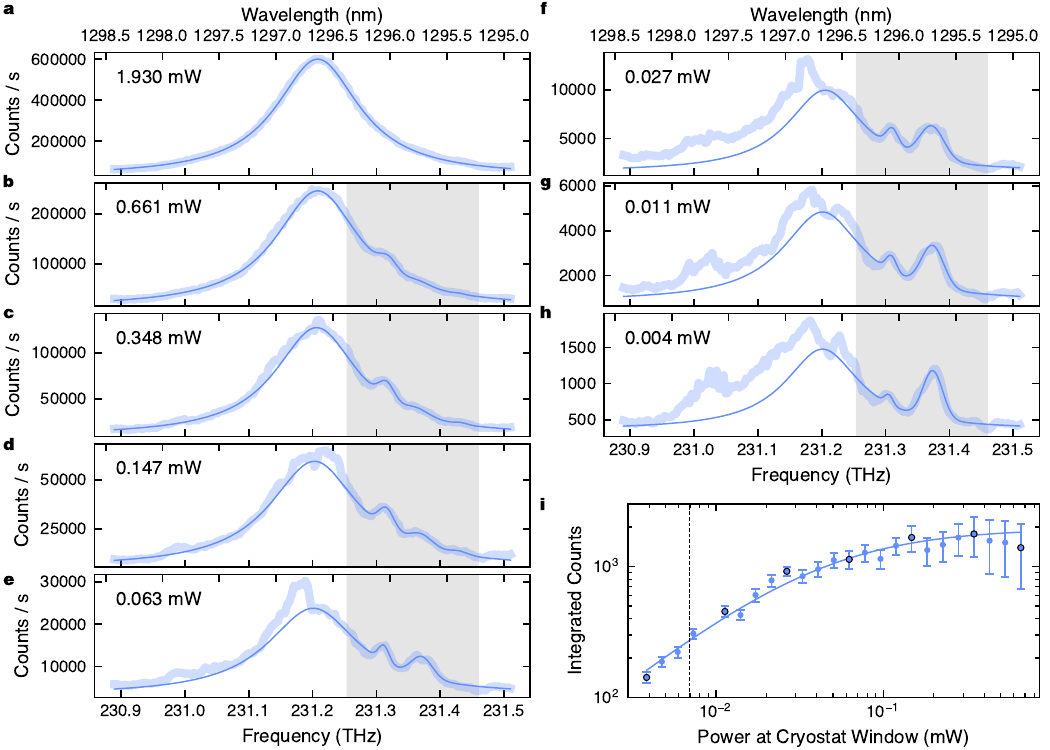}
    \caption[Saturation of the quantum dot emission]{\textbf{a}-\textbf{h} Emission spectra for QDs coupled to the microdisk resonator, collected at the right output port (cf.~Fig.~1a) under a 9~T applied magnetic field, for excitation powers (at the cryostat window) decreasing from a maximum of 1.93~mW to a minimum of 4~\textmu W. The data (thick faded line) at the maximum power in \textbf{a} only resolves the line shape of the cavity, so the fit (thin dark line) includes the cavity response alone. At all other powers, in panels \textbf{b}-\textbf{h}, two QD emission lines are fit inside a truncated spectral window (shaded grey region) after the cavity response is adjusted for the power. The QD emission line at higher frequency (to the right) is that of interest in this work. \textbf{i} Total emission intensity (markers with 95\% confidence interval error bars) for the QD of interest as a function of excitation power, and a saturation curve fit (blue line). The markers outlined in black correspond to the spectra shown in panels \textbf{b}-\textbf{h}, and the dashed black line highlights the excitation power used to produce the data shown in Fig.~2a of the main text.}
    \label{fig:saturation}
\end{figure}

Following the procedure for measuring QD emission spectra described in Sec.~V of the main text, with a constant 9~T magnetic field applied and collecting only from the right port (shown in Fig.~1a), we swept the power of the excitation laser from 4~\textmu W to 1.93 mW (at the cryostat window) to probe the emission saturation of the QD of interest. Representative example spectra at eight different powers are shown across Figs.~\ref{fig:saturation}a-h.
At the highest power (Fig.~\ref{fig:saturation}a), all QD emission lines in the spectral window are saturated, joining together to fill the cavity resonance. This allows us to fit the cavity spectrum using a Lorentzian line shape, following Eq.~\ref{eq:cavity}. We find the cavity to be centred at $f_\mathrm{cav} = 231.2083\pm0.0006$~THz with a FWHM of $\gamma_\mathrm{cav}=126\pm7$~GHz, where the errors are 95\% confidence intervals calculated by expanding the standard errors returned by the fit using the t-distribution. Conversely, in Figs.~\ref{fig:saturation}b-h, emission into the cavity has not fully overtaken the QD emission lines, so we can resolve and fit them in order to evaluate their overall emission intensity. To isolate the QD emission lines, we first perform cavity subtraction on each individual spectrum by using the result from Fig.~\ref{fig:saturation}a to find the background underneath the QD emission (see Sec.~\ref{sec:spectra-fits} for more details). Then, we fit all emission lines within a truncated spectral window around the QD of interest (shaded grey region), modelling each using a Gaussian line shape,
\begin{equation} \label{eq:qd_emission_line}
    I_\mathrm{qd}(f) = I_\mathrm{qd,0}\exp\left(-4\ln{(2)}\left(\frac{f-f_\mathrm{qd}}{\gamma_\mathrm{qd}}\right)^2\right),
\end{equation}
where $I_\mathrm{qd,0}$ is the peak intensity, $\gamma_\mathrm{qd}$ is the FWHM of the QD emission line, and $f_\mathrm{qd}$ is its centre frequency. The Gaussian line shape was found to yield a better fit to the data, likely due to the broadening induced by both spectral diffusion \cite{Wakileh:26} and the Gaussian-shaped tunable bandpass filter (0.1~nm bandwidth) used during measurement. For these spectra, we resolve and fit the emission from one neighbouring QD (lower frequency, left peak) in addition to the QD of interest (higher frequency, right peak). For each Gaussian line fit, we integrate over all frequency to extract the overall emission intensity, which is shown as a function of excitation power in Fig.~\ref{fig:saturation}i. Here, the error bars are 95\% confidence intervals calculated by propagating errors from the fit to the integration. We fit a saturation curve to this data,
\begin{equation} \label{eq:saturation}
    I_\mathrm{tot}(P) = I_\mathrm{tot}(\infty)\frac{P}{P+P_\mathrm{sat}},
\end{equation}
where $I_\mathrm{tot}$ is the total intensity as a function of excitation power $P$, and $P_\mathrm{sat}$ is the saturation power for the QD. Ultimately, we find $P_\mathrm{sat} = 43\pm5$~\textmu W, which is a factor of 6 greater than the excitation power used during the measurements shown in Fig.~2a of the main text (7~\textmu W, dashed black line in Fig.~\ref{fig:saturation}i). 

\section{Fitting Quantum Dot Emission Spectra} \label{sec:spectra-fits}
\begin{figure}[htb]
    \includegraphics{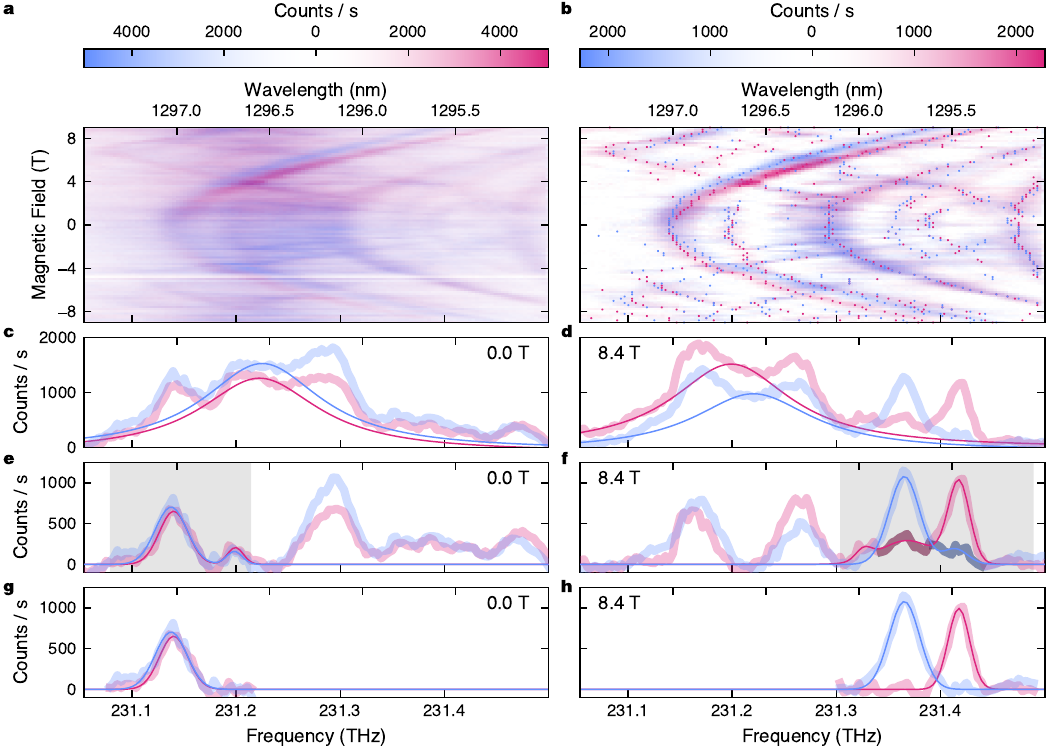}
    \caption[Analysis of QD emission spectra for all applied magnetic fields]{\textbf{a} Raw emission spectra for all applied magnetic fields, where those collected from the left (pink) and right (blue) output ports are overlaid. \textbf{b} Emission spectra after cavity subtraction. Circular markers denote peaks identified for subsequent QD emission line fitting. \textbf{c}, \textbf{d} Emission spectra recorded at 0~T and 8.4~T, respectively, where thick, faded lines correspond to the raw data while thin, dark lines show a fit of background emission from poorly coupled QDs into the cavity resonance. \textbf{e}, \textbf{f} Emission spectra for 0~T and 8.4~T, respectively, after the cavity background is subtracted. QD emission lines identified within a truncated window around the QD of interest (shaded grey regions) are fit (thin, dark lines) to the data (thick, faded lines). In \textbf{f}, emission from an overlapping neighbouring QD, considered further in Fig.~\ref{fig:nearby_dots}, is highlighted by darkening the curve. \textbf{h} Emission spectra at 0~T and 8.4~T, respectively, for the QD of interest alone after completing the fitting.}
    \label{fig:spectra_fitting}
\end{figure}

The raw data for the emission spectra taken around the QD of interest as a function of the applied magnetic field, including emission from noisy background QDs into the cavity resonance of the microdisk, is displayed in Fig.~\ref{fig:spectra_fitting}a.
Here, it is evident that the cavity spans the entire spectral window, and due to the high QD density on this sample, there is a significant amount of background emission (purple haze across the image). To extract accurate fits of the quantum dot emission lines within the spectral window, and better identify the presence of neighbouring dots around the QD of interest, we subtract the cavity background first, resulting in Fig.~\ref{fig:spectra_fitting}b. Cavity subtraction was performed for each individual spectrum (each propagation direction at each magnetic field) by fitting the Lorentzian line shape of the cavity (Eq.~\ref{eq:cavity}) to the data while maintaining tight bounds on the FWHM and resonance frequency of the cavity, centred around the results extracted from the high-power spectrum of Fig.~\ref{fig:saturation}a (see Sec.~\ref{sec:saturation} for more details). The first fit is poor because it tends upward in an attempt to include the QD emission lines in the cavity. Thus, we repeat the fit twice more, each time removing data points that yield positive residuals greater than 1\% of the overall maximum counts. This ensures that the data points where the QD emission lines are prominent are not included in the fit of the cavity background, with a buffer for noise.

Examples of the cavity fit (thin dark red and blue lines) are shown in Figs.~\ref{fig:spectra_fitting}c and d for 0~T and 8.4~T magnetic fields, the same examples chosen for Figs.~2c and d in the main text. It is evident that the cavity background, and corresponding fits, differ slightly for the two output ports and at different magnetic fields. While spectra from opposite ports were taken in quick succession, we expect that either partial directionality from the poorly-coupled QDs that fill the cavity resonance, or hysteresis effects from the tunable filter used when recording the spectra, introduce these slight asymmetries. In contrast, the magnetic field sweep was performed over the course of multiple weeks, meaning that small variations in the calibration of the tunable filter are likely the cause of the shift apparent between panels c with d. However, when comparing these panels to those after subtraction, Figs.~\ref{fig:spectra_fitting}e and f, it is clear that the cavity fits accurately capture the background emission without distorting the QD emission lines.

Following cavity subtraction, we apply peak finding across each individual spectrum to identify all of the QD emission lines (see markers in Fig.~\ref{fig:spectra_fitting}b). These peaks inform the subsequent fits on the number of QD lines to include as well as the initial guesses for the centre frequency of each QD. Since there are many neighbouring dots within each full spectrum, we truncate the spectral window around the QD of interest (shaded grey regions in Figs.~\ref{fig:spectra_fitting}e and f). Subsequently, emission from the identified QDs within the truncated window is fit using Gaussian line shapes (Eq.~\ref{eq:qd_emission_line}), and the QD of interest is fully isolated (Figs.~\ref{fig:spectra_fitting}g and h) for further analysis on Zeeman splitting (Fig.~2b) and directional emission (Fig.~3, and described further in Sec.~\ref{sec:direction}).

\begin{figure}[p]
    \includegraphics{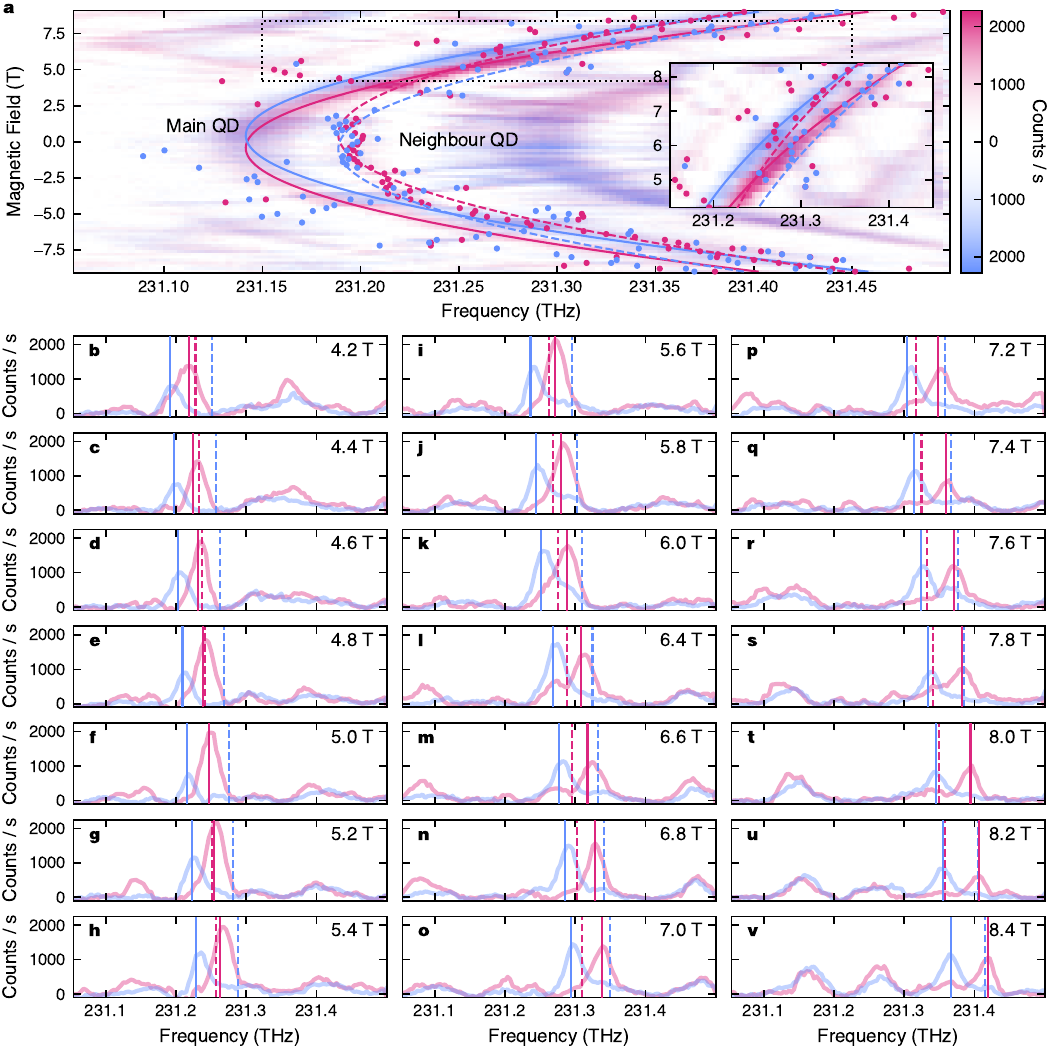}
    \caption[Exemplary neighbouring QD identified in the emission spectra]{\textbf{a} QD emission spectra, after cavity subtraction (as in Fig.~\ref{fig:spectra_fitting}b), for all applied magnetic fields, where those collected from the left (pink) and right (blue) output ports are overlaid. Circular markers denote centre frequencies extracted from fits of nearby QDs around the QD of interest. Solid pink and blue lines highlight the magnetic field tuning of the main QD, while the dashed lines indicate a likely trajectory for the two dipole transitions of a neighbour QD. The neighbour emits at a higher frequency than the main QD at 0~T (to the right), then tunes into the main QD as the magnetic field increases in amplitude for both polarities. The inset, corresponding to the dotted box region, provides an expanded view of these trajectories over the 4.2~T to 8.4~T magnetic field range, in correspondence with the other panels. \textbf{b}-\textbf{v} For all applied magnetic fields from 4.2~T to 8.4~T, in steps of 0.2~T, individual cavity-subtracted spectra (thick, faded curves) are shown for the QD emission in each direction (pink for left and blue for right). The solid (dashed) vertical lines indicate the centre frequencies of the two main (neighbour) QD dipoles at each field.}
    \label{fig:nearby_dots}
\end{figure}

Looking at the cavity-subtracted spectrum for an applied magnetic field of 8.4~T (Fig.~\ref{fig:spectra_fitting}f) alone, it appears that emission lines underneath those for the QD of interest (darkened regions of the curve), in the opposite direction, may be indicative of a reduced directionality. However, when this spectrum is considered in the context of the full magnetic field sweep, it is evident that these emission lines arise from a neighbouring QD rather than the QD of interest. To illustrate this, we again plot the emission spectra at all magnetic fields in Fig.~\ref{fig:nearby_dots}a,
yet now annotate circular markers (pink and blue for left and right ports, respectively) that show the centre frequencies extracted when fitting nearby QDs around that of interest. Near 0~T, a neighbour QD clearly emits at a slightly higher frequency (to the right) than the QD of interest. While the emission from this neighbour QD is far less bright, and thus not resolvable in each direction at all magnetic fields, it appears frequently enough that we can trace a likely trajectory for the dipoles of this QD as they tune (dashed pink and blue lines). Following this trajectory, we find that the neighbour QD tends to overlap closely with the main QD (solid pink and blue lines) at high-amplitude magnetic fields of both polarities.

Through a closer examination of the individual spectra taken in steps of 0.2~T from 4.2~T to 8.4~T (shown in Figs.~\ref{fig:nearby_dots}b-v), it is clear that the trajectory traced in Fig.~\ref{fig:nearby_dots}a accurately describes the tuning of the neighbour QD. From 4.2~T to 5.0~T (Figs.~\ref{fig:nearby_dots}b-f), detected emission from one dipole of the neighbour QD (centred at the dashed lines) approaches one dipole of the main QD (centred at the solid lines) from a higher frequency (right side), causing the intensity measured in the left port (pink) to increase. As the magnetic field continues to increase toward 5.8~T (Figs.~\ref{fig:nearby_dots}g-j), the second neighbour QD dipole (right, blue) approaches while the first dipole (left, red) continues to overlap with the main QD, and can be seen emitting between the two transition dipoles of the main QD at 6.0~T (Fig.~\ref{fig:nearby_dots}k), consistent with the trajectory shown in the inset of Fig.~\ref{fig:nearby_dots}a. Once the magnetic field reaches 6.6~T (Fig.~\ref{fig:nearby_dots}m), the trajectories of the two dipoles of the neighbour QD are closely overlapping with those of the main QD, and this trend continues as the field approaches 8.4~T (Fig.~\ref{fig:nearby_dots}v). While this particular neighbour QD overlaps at the most magnetic field values, this form of analysis was repeated for all other neighbour QDs to ensure the analysis of the main QD is consistent over the entire dataset.

\section{Fitting Quantum Dot Emission Lifetimes} \label{sec:lifetime}
Lifetime measurement data was fit using a convolution of a Gaussian instrument response function with a decaying exponential, given by,
\begin{equation}
L(t,\tau,A,\sigma,t_a) =A\exp{\left[\frac{\sigma^2}{2\tau^2}-\frac{(t-t_a)}{\tau}\right]}\left[1-\operatorname{erf}\left(\frac{-(t-t_a)+\sigma^2/\tau}{\sqrt{2}\sigma}\right)\right],
\end{equation}
where $\sigma$ is the temporal width of the Gaussian instrument response function, $\tau$ is the emission lifetime, $t_a$ is the arrival time of the pulse, and $A$ is the amplitude. For each lifetime, at each applied magnetic field, a bi-exponential decay,
\begin{equation}
    L(t,\tau_1,A_1,\sigma,t_a) + L(t,\tau_2,A_2,\sigma,t_a) + L_0,
\end{equation}
was used for the initial fit, where $L_0$ is a baseline that accounts for background counts. If the fit errors on either amplitude parameter exceeded 100\%, it indicated that the data was overfit. Subsequently, the fit function was updated to a mono-exponential. This was only the case for magnetic fields ranging from -9.0~T to -7.5~T. Once lifetime values were extracted for each magnetic field, a Lorentzian fit was applied to the data following \cite{Barbour:11},
\begin{equation} \label{eq:lifetime-purcell}
    \tau_\mathrm{QD}(\Delta) = \tau_\mathrm{QD, 0} - \frac{A_\mathrm{dip}\gamma_\mathrm{cav}^2}{\gamma_\mathrm{cav}^2 + \Delta^2},
\end{equation}
where $\tau_\mathrm{QD, 0}$ is the far-off-resonant lifetime of the QD, $A_\mathrm{dip}$ is the amplitude corresponding to the decrease in lifetime from Purcell enhancement, $\gamma_\mathrm{cav}$ is the FWHM of the cavity resonance (obtained as described in Sec.~\ref{sec:saturation}), and $\Delta$ is the cavity-QD detuning. Previous measurements of detuning as a function of magnetic field (see Fig.~2a and b) were used to evaluate Eq.~\ref{eq:lifetime-purcell}, while $\tau_\mathrm{QD, 0}$ and $A_\mathrm{dip}$ were fit parameters. The quadratic dependence of the detuning on the magnetic field causes the QD to tune into resonance with the cavity at both positive and negative fields, thus leading to two distinct dips as opposed to one.

\section{Extracting Directionality and Directional Contrast}\label{sec:direction}
In the literature, several figures of merit are used to quantify directional QD emission, especially for experimental results. A common choice is the directional contrast \cite{Coles:16, Martin:25},
\begin{equation} \label{eq:contrast}
    C_{\mathrm{L}/\mathrm{R}} = \frac{I_{+,\mathrm{L}/\mathrm{R}}-I_{-,\mathrm{L}/\mathrm{R}}}{I_{+,\mathrm{L}/\mathrm{R}}+I_{-,\mathrm{L}/\mathrm{R}}},
\end{equation}
where $I_{+,\mathrm{L}/\mathrm{R}}$ and $I_{-,\mathrm{L}/\mathrm{R}}$ refer to the emission intensity from the $\sigma^+$- and $\sigma^-$-polarized QD transition dipoles collected from the left (L) or right (R) ports, respectively. Since the contrast is evaluated for each output port independently, it is insensitive to differences in collection efficiency between the two directions. However, a more intuitive metric for directional emission is the directionality \cite{Schrinski:22}, given by,
\begin{equation} \label{eq:directionality}
    D_\pm = \frac{I_{\pm,\mathrm{L}}-I_{\pm,\mathrm{R}}}{I_{\pm,\mathrm{L}}+I_{\pm,\mathrm{R}}},
\end{equation}
where the emission intensity of one transition dipole ($\sigma^+$ or $\sigma^-$) is compared across the two output ports. This definition allows one to directly quantify the fraction of emission directed into the preferred output, yet is sensitive to an imbalance in collection efficiencies as we describe further below.

To evaluate either the directional contrast or directionality from the data, we must calculate four intensities: $I_{+,\mathrm{L}}$, $I_{+,\mathrm{R}}$, $I_{-,\mathrm{L}}$, $I_{-,\mathrm{R}}$. Following the fitting procedure described in Sec.~\ref{sec:spectra-fits} (examples of which are shown in Fig.~\ref{fig:spectra_fitting}), we are able to extract the centre frequency and linewidth for each transition dipole of the QD of interest at each magnetic field. Then, the respective intensities for each dipole in each output port are calculated by summing the counts within one FWHM of the corresponding centre frequency in the data for the corresponding direction. To calculate 95\% confidence intervals for directional contrast and directionality at each field, we build distributions of the FWHMs and centre frequencies that are propagated from the QD emission line fit errors. We then sample from these distributions in 5000 trials, each time recalculating the directional contrast and directionality. In the end, we have a distribution for each measure from which the confidence intervals are extracted. Note that these calculations are conducted specifically on the data after background emission from the cavity and nearby QDs is subtracted (see Sec.~\ref{sec:spectra-fits}).

While the collection efficiencies for the measured QD emission spectra are similar across the two output ports, they are not identical, and this skews the directionality results particularly at magnetic fields approaching 0~T before the two transition dipoles have fully split in emission frequency. Thus, in line with previous works \cite{Coles:16, Martin:25, Mehrabad:20}, we evaluate only the directional contrast for the QD of interest at all magnetic fields (see Fig.~3), and evaluate the directionality only at magnetic fields with an absolute value greater than 7.5~T (indicated by the dark teal regions in Fig.~3). To fit the magnetic field dependence of the directional contrast, in each respective direction, we construct a model of two Gaussian emission lines that share a common FWHM and follow the Zeeman splitting extracted from the data (as shown in Fig.~2b). The common FWHM is a fit parameter, alongside the maximum directional contrast, which directly dictates the relative amplitudes of the two Gaussian emission lines. These parameters fully define the two emission lines, from which we evaluate the directional contrast in the same way that is conducted for the measured data, by summing the counts within the FWHM centred on each line. Altogether, the model describes a fit function with a directional contrast that tends to 0 at 0~T, yet increases in magnitude to a maximum at high-amplitude magnetic fields when the two transition dipoles of the QD are fully split. From the definition in Eq.~\ref{eq:contrast}, it is clear that the dipole which emits preferentially to the left has a positive contrast, while that which emits to the right has a negative one. Thus, to achieve the final reported value of $C=0.980~[0.939, 0.994]$, we take the absolute value of the contrasts extracted from the fits for each direction and average them. The confidence intervals are calculated by propagating the standard errors extracted from each fit, then expanding to a 95\% confidence interval using the t-distribution. For the directionality, the procedure is similar, yet we assume that the two Gaussian emission lines are fully split at the magnetic fields considered (absolute value greater than 7.5~T). The final result is $D=0.986~[0.905, 0.998]$, where the confidence interval is broader due to the truncation of the magnetic field range.

We now return to the comparison between directional contrast and directionality to provide context for the discussion of Fig.~4 in the main text. In Fig.~4, we compare experimentally measured directional contrasts with theoretical directionality thresholds for a range of quantum information protocols. While the nomenclature is distinct, this is in fact a comparison of equivalent measures when the total emission intensities from each transition dipole of the QD are balanced. Specifically, the protocols we highlight each require some threshold directionality on-chip, at the QD used to form the quantum photonic device. Inside the device, directional contrast and directionality are equivalent because outcoupling has not occurred, and thus no coupling asymmetries are introduced. To see this explicitly, consider the four intensities that are used in Eqs.~\ref{eq:contrast}, \ref{eq:directionality}. When outcoupling efficiency is balanced in the left and right output ports, we can define the total intensity as $I_\mathrm{tot} \equiv I_{+,\mathrm{L}} + I_{-,\mathrm{L}} = I_{+,\mathrm{R}} + I_{-,\mathrm{R}}$. If the overall intensity emitted from each transition dipole is also balanced, which is the case for QDs chirally coupled to waveguides or those where the cavity-dipole detuning is approximately equivalent (as in this work at large magnetic fields where both transitions are far off-resonant), then $I_\mathrm{tot} = I_{+,\mathrm{L}} + I_{+,\mathrm{R}} = I_{-,\mathrm{L}} + I_{-,\mathrm{R}}$ as well. Substituting these definitions into Eq.~\ref{eq:directionality}, we find,
\begin{align} \label{eq:dir-math}
    D &= \frac{1}{2}\left(\left|D_+\right| + |D_-|\right),\nonumber\\
    D &= \frac{1}{2}\left(\frac{\left|I_{+,\mathrm{L}} - I_{+,\mathrm{R}}\right|}{I_\mathrm{tot}} + \frac{\left|I_{-,\mathrm{L}} - I_{-,\mathrm{R}}\right|}{I_\mathrm{tot}}\right),\nonumber \\
    D &= \frac{1}{2I_\mathrm{tot}}\left(I_{+,\mathrm{R}} - I_{+,\mathrm{L}} + I_{-,\mathrm{L}} - I_{-,\mathrm{R}}\right),
\end{align}
where in the last step we assume that the $\sigma^+$ dipole emits preferentially to the right ($I_{+,\mathrm{R}} > I_{+,\mathrm{L}}$) and the $\sigma_-$ dipole to the left ($I_{-,\mathrm{L}} > I_{-,\mathrm{R}}$), as is the case in this work, but the final result is the same in the opposite case. Rearranging Eq.~\ref{eq:dir-math}, then substituting Eq.~\ref{eq:contrast},
\begin{align} \label{eq:DisC}
    D &= \frac{1}{2}\left(\frac{I_{-,\mathrm{L}}-I_{+,\mathrm{L}}}{I_\mathrm{tot}} + \frac{I_{+,\mathrm{R}}-I_{-,\mathrm{R}}}{I_\mathrm{tot}}\right),\nonumber\\
    D &= \frac{1}{2}\left(\frac{\left|I_{+,\mathrm{L}}-I_{-,\mathrm{L}}\right|}{I_\mathrm{tot}} + \frac{\left|I_{+,\mathrm{R}}-I_{-,\mathrm{R}}\right|}{I_\mathrm{tot}}\right),\nonumber\\
    D &= \frac{1}{2}\left(C_\mathrm{L} + C_\mathrm{R}\right),\nonumber\\
    D &= C,
\end{align}
we find the equivalence of directionality and directional contrast under the aforementioned assumptions.

Separately, although Ref.~\cite{Sollner:15} does not explicitly report the directional contrast nor directionality, these values can be extracted from their reported directional factors, defined as $F_{\mathrm{dir},\mathrm{L}/\mathrm{R}}=\frac{I_{+,\mathrm{L}/-,\mathrm{R}}}{I_{+,\mathrm{L}/\mathrm{R}}+I_{-,\mathrm{L}/\mathrm{R}}}$, with an average value of $F_\mathrm{dir} =(F_{\mathrm{dir},L}+F_{\mathrm{dir},R})/2 = 0.900\pm0.013$. Since the authors do not report individual values for left and right directions, we assume $F_{\mathrm{dir},\mathrm{L}} \approx F_{\mathrm{dir},\mathrm{R}}$, which is consistent with the data shown (Fig.~S4 in Ref.~\cite{Sollner:15}). Rearranging Eq.~\ref{eq:contrast} for the left port, and substituting the directional factor definition, we find,
\begin{align}
    C_\mathrm{L} &= \frac{I_{+,\mathrm{L}}}{I_{+,\mathrm{L}} + I_{-,\mathrm{L}}} - \frac{I_{-,\mathrm{L}}}{I_{+,\mathrm{L}} + I_{-,\mathrm{L}}},\nonumber\\
    C_\mathrm{L} &= F_\mathrm{dir,L} - \frac{I_{-,\mathrm{L}} + I_{+,\mathrm{L}} - I_{+,\mathrm{L}}}{I_{+,\mathrm{L}} + I_{-,\mathrm{L}}},\nonumber\\
    C_\mathrm{L} &= F_\mathrm{dir,L} - \frac{I_{+,\mathrm{L}} - I_{-,\mathrm{L}}}{I_{+,\mathrm{L}} + I_{-,\mathrm{L}}} + \frac{I_{+,\mathrm{L}}}{I_{+,\mathrm{L}} + I_{-,\mathrm{L}}},\nonumber\\
    C_\mathrm{L} &= 2F_\mathrm{dir,L} - 1.
\end{align}
The procedure is the same for the right port, and allows us to estimate the average directional contrast as $C=0.800\pm0.026$.

\section{Measured Directional and Symmetric Emission from Other Quantum Dots} \label{sec:otherdots}
In addition to the QD of interest discussed extensively in the main text, we characterized four other O-band QDs, two of which were found to emit directionally and two symmetrically. The other directional QDs were found similarly in microdisk structures, coupled to respective cavity resonances, while the symmetric QDs were each coupled to a waveguide. The measurement procedure for these dots is analogous to that described in Sec.~V of the main text for the QD of interest, however, the magnetic field ranges vary between the dots, the excitation powers are slightly increased ($\sim 15$~\textmu W at the cryostat entrance window), and the spectra were measured on the spectrometer (0.04~nm-resolution grating and a liquid nitrogen-cooled InGaAs linear array detector) rather than the combination of a tunable filter with superconducting nanowire single-photon detectors.

\begin{figure}[p]
    \includegraphics{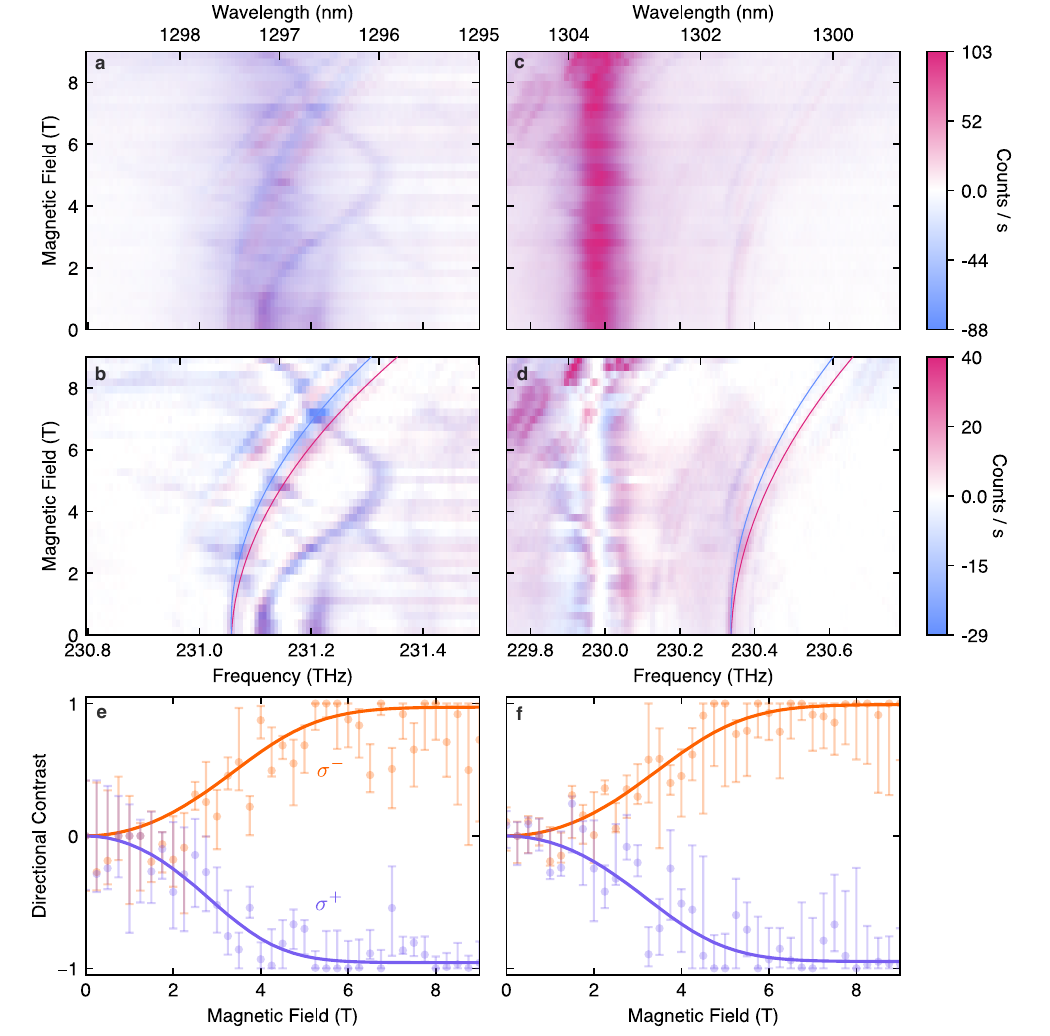}
    \caption[Additional directional QDs coupled to a microdisk]{\textbf{a,b} Emission spectra recorded from the left (pink) and right (blue) output ports, as the magnetic field is varied from 0~T to 9~T in 0.25~T increments, before (\textbf{a}) and after (\textbf{b}) cavity subtraction, respectively. In \textbf{b}, a fit (curves) to the magnetic field tuning of each transition dipole of the QD is shown. \textbf{c,d} Analogous emission spectra before and after cavity subtraction, respectively, for a second QD. \textbf{e,f} Directional contrast for both the $\sigma^+$ (purple) and $\sigma^-$ (orange) transition dipoles, shown with markers and corresponding 95\% confidence intervals (error bars) as a function of the applied magnetic field, for each additional directional QD, respectively. Fits to the directional contrast for each transition dipole of each QD are overlaid (curves).}
    \label{fig:otherdir}
\end{figure}

For the directional QDs, the measured emission spectra are shown in Figs.~\ref{fig:otherdir}a and c,
respectively. We used the data gathered when characterizing the cavity resonances for the analysis in Sec.~\ref{sec:Qfactors} (high-power excitation) to perform cavity subtraction (following the same procedure as described in Sec.~\ref{sec:spectra-fits}), resulting in Figs.~\ref{fig:otherdir}b and d, respectively. Then, we fit Gaussian emission lines for both the main and neighbouring QDs at each magnetic field. The analysis of directional contrast was conducted using the same procedure as that described in Sec.~\ref{sec:direction} (see results in Figs.~\ref{fig:otherdir}e,f). Ultimately, we calculate $C = 0.965~[0.917, 0.986]$ and $C = 0.970~[0.420, 0.999]$, respectively. Returning to Figs.~\ref{fig:otherdir}a-d, it appears that one of the neighbouring dots in each set of emission spectra also shows highly directional emission, however, these dots were poorly resolved at magnetic fields approaching 0~T and thus were not considered for further analysis.

\begin{figure}[p]
    \includegraphics{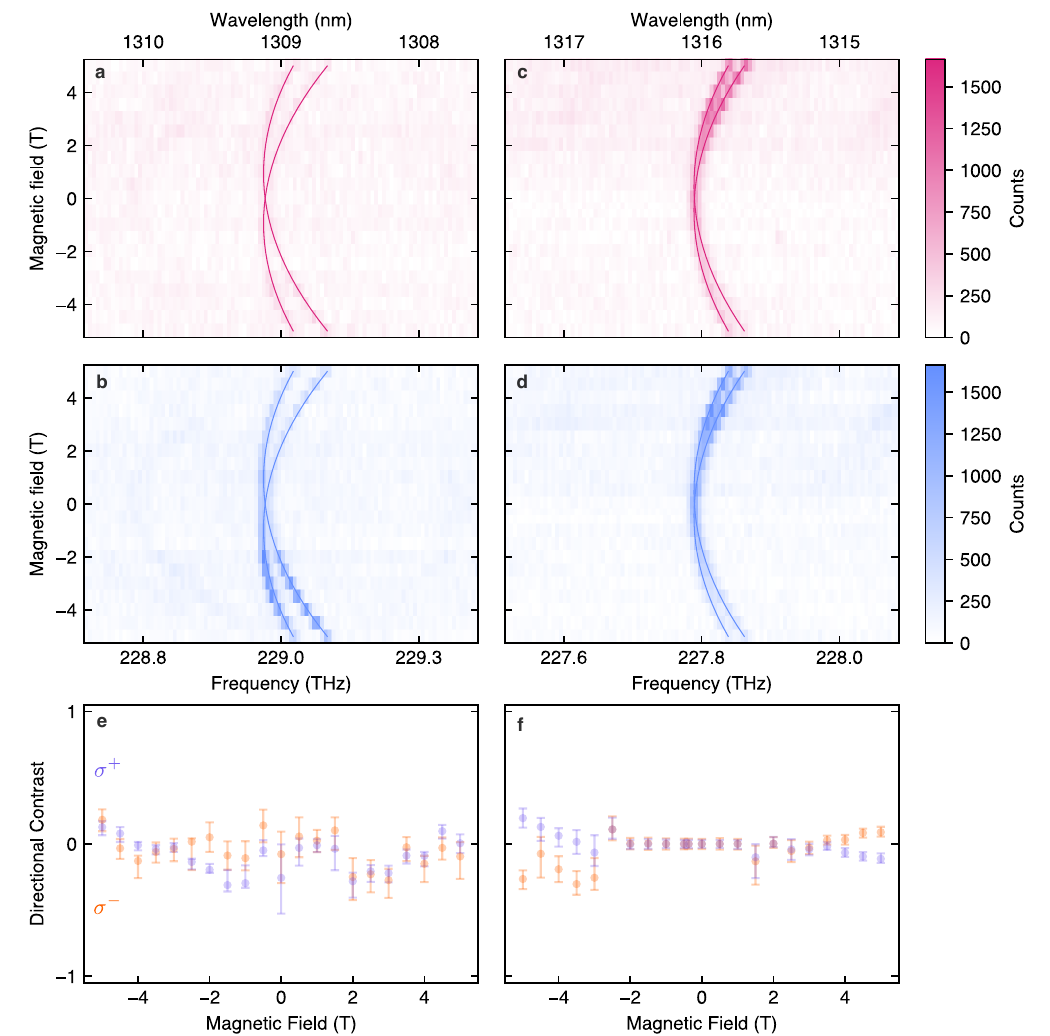}
    \caption[Low directional contrast for QDs symmetrically coupled to a waveguide]{\textbf{a,b} Emission spectra of one QD recorded from the left (pink) and right (blue) output ports, respectively, as the magnetic field is varied from -5~T to 5~T in 0.5~T increments. Both branches of the Zeeman splitting (see overlaid fit curves), from the two QD transition dipoles, appear in both ports. \textbf{c,d} Emission spectra for a second QD, recorded from the left (pink) and right (blue) ports, respectively; both Zeeman branches again appear in both ports, and a fit of the magnetic field tuning is again overlaid (curves). \textbf{e,f} Directional contrasts (markers), with corresponding 95\% confidence intervals (error bars), for the $\sigma^+$ (purple) and $\sigma^-$ (orange) dipoles of each QD, respectively, as a function of magnetic field. Both datasets show no apparent directional trend across the full range of magnetic fields.}
    \label{fig:symm}
\end{figure}

Since the symmetric QDs were each coupled to a waveguide, there was no cavity subtraction performed on the measured emission spectra (see Figs.~\ref{fig:symm}a-d, where data measured from the two output ports is now separated for visual clarity).
Also, the absence of neighbouring QDs allowed us to fit each QD of interest alone. In doing so, two Gaussian QD emission lines, for the two respective transition dipoles, were fit for each output port at all magnetic fields (except those where only a single Gaussian could be resolved, -2.5~T to 2~T for the QD shown in Figs.~\ref{fig:symm}c,d). From the fits to the spectra, the directional contrast was calculated at each magnetic field again following the same procedure as explained in Sec.~\ref{sec:direction}, resulting in the data shown in Figs.~\ref{fig:symm}e,f. From this data, neither QD exhibits a trend of directional emission. Thus, rather than using the model from Sec.~\ref{sec:direction}, we instead fit a line to the data for each transition dipole, resulting in $C=0.053~[0.022, 0.088]$ and $C=0.010~[0.000, 0.029]$ for Figs.~\ref{fig:symm}e and f, respectively.

\bibliography{references}